\documentclass[12pt,preprint]{aastex}

\shorttitle{HiRes UHECR Composition Study}
\shortauthors{The HiRes Collaboration}

\begin{document}

\title{A Study of the Composition of Ultra High Energy Cosmic Rays Using the High Resolution Fly's Eye}

\author{
R.U. Abbasi\altaffilmark{1}, 
T. Abu-Zayyad\altaffilmark{1}, 
G. Archbold\altaffilmark{1,2}, 
R. Atkins\altaffilmark{1},
J. Bellido\altaffilmark{3}, 
K. Belov\altaffilmark{1}, 
J.W. Belz\altaffilmark{4}, 
S. BenZvi\altaffilmark{5}, 
D.R. Bergman\altaffilmark{6}, 
J. Boyer\altaffilmark{5}, 
G.W. Burt\altaffilmark{1}, 
Z. Cao\altaffilmark{1}, 
R. Clay\altaffilmark{3}, 
B.M. Connolly\altaffilmark{5}, 
B. Dawson\altaffilmark{3}, 
W. Deng\altaffilmark{1}, 
Y. Fedorova \altaffilmark{1}, 
J. Findlay\altaffilmark{1},
C.B. Finley\altaffilmark{5}, 
W.F. Hanlon\altaffilmark{1}, 
G.A. Hughes\altaffilmark{6}, 
P. Huntemeyer \altaffilmark{1}, 
C.C.H. Jui\altaffilmark{1}, 
K. Kim\altaffilmark{1}, 
M.A. Kirn\altaffilmark{4}, 
B. Knapp\altaffilmark{5}, 
E.C. Loh\altaffilmark{1}, 
M.M. Maetas\altaffilmark{1}, 
K. Martens\altaffilmark{1}, 
G. Martin\altaffilmark{7}, 
N. Manago\altaffilmark{8}, 
E.J. Mannel\altaffilmark{5}, 
J.A.J. Matthews\altaffilmark{7}, 
J.N. Matthews\altaffilmark{1}, 
A. O'Neill\altaffilmark{5}, 
L. Perera\altaffilmark{6}, 
K. Reil\altaffilmark{1}, 
R. Riehle\altaffilmark{1}, 
M.D. Roberts\altaffilmark{7}, 
M. Sasaki\altaffilmark{8}, 
M. Seman\altaffilmark{5}, 
S.R. Schnetzer\altaffilmark{6}, 
K. Simpson\altaffilmark{3}, 
J.D. Smith\altaffilmark{1}, 
R. Snow\altaffilmark{1}, 
P. Sokolsky\altaffilmark{1}, 
C. Song\altaffilmark{5}, 
R.W. Springer\altaffilmark{1}, 
B.T. Stokes\altaffilmark{1}, 
J.R. Thomas\altaffilmark{1},
S.B. Thomas\altaffilmark{1}, 
G.B. Thomson\altaffilmark{6}, 
S. Westerhoff\altaffilmark{5}, 
L.R. Wiencke\altaffilmark{1}, and 
A. Zech\altaffilmark{6}}

\affil{The High Resolution Fly's Eye Collaboration}

\altaffiltext{1}{University of Utah, Department of Physics and 
High Energy Astrophysics Institute,
Salt Lake City, UT 84112, USA}

\altaffiltext{2}{present address: Lawrence Livermore National Laboratory, Livermore, CA 94551, USA: archbold1@llnl.gov}

\altaffiltext{3}{University of Adelaide, Department of Physics,
Adelaide, SA 5005, Australia}

\altaffiltext{4}{University of Montana, Department of Physics and Astronomy,
Missoula, MT 59812, USA}

\altaffiltext{5}{Columbia University, Department of Physics, 
Nevis Laboratories, Irvington, NY 10027, USA}

\altaffiltext{6}{Rutgers University - The State University of 
New Jersey, Department of Physics and Astronomy, Piscataway,
NJ 08854, USA}

\altaffiltext{7}{University of New Mexico, Department of Physics 
and Astronomy, Albuquerque, NM 87131, USA}

\altaffiltext{8}{University of Tokyo, Institute for Cosmic 
Ray Research, Kashiwa City, Chiba 277-8582, Japan}

\begin{abstract}
The composition of Ultra High Energy Cosmic Rays (UHECR) is measured
with the High Resolution Fly's Eye cosmic ray observatory (HiRes) data
using the X$_{max}$ technique.  Data were collected in stereo between 1999 November and 2001 September.  The data are reconstructed with well-determined geometry.  Measurements of the atmospheric transmission are incorporated in the reconstruction.  The detector resolution is found to be 30 g cm$^{-2}$ in X$_{max}$ and 13\% in Energy.  The X$_{max}$ elongation rate between $10^{18.0}$ eV and $10^{19.4}$ eV is measured to be 54.5 $\pm$ 6.5 (stat) $\pm$ 4.5 (sys) g cm$^{-2}$ per decade.  This is compared to predictions using the QGSJet01 and SIBYLL 2.1 hadronic interaction models for both protons and iron nuclei.  CORSIKA-generated Extensive Air Showers (EAS) are incorporated directly into a detailed detector Monte Carlo program. The elongation rate and the X$_{max}$ distribution widths are consistent with a constant or slowly changing and predominantly light composition.  A simple model containing only protons and iron nuclei is compared to QGSJet and SIBYLL.  The best agreement between the model and the data is at 80\% protons for QGSJet and 60\% protons for SIBYLL.

\end{abstract}


\section{INTRODUCTION}
\label{introduction}

The cosmic ray (CR) spectrum follows a power law which exhibits several interesting features.  A break (known as the knee) between a power law with index -2.7 to index -3.0 occurs near $10^{15}$ eV.  A similar break (the second knee) from index -3.0 to -3.3 has been reported to occur near $10^{17.7}$ eV \citep{abuaj, 38, bird95, prav99, prav03, nag84, nag91}, followed by a rise to an index of -2.7 (the ankle) near $10^{18.6}$ eV.  The spectrum appears to continue until $10^{19.8}$ eV.

Changes in spectral index at various energies could be produced by a gradual decrease in efficiency with energy of a source (e.g., galactic supernovae), leakage out of an area of magnetic confinement (the galactic leaky box model), the appearance of flux from new sources that begin to dominate at higher energies (extra-galactic cosmic rays), and thresholds of inelastic interactions between CR 
protons and the cosmic microwave background radiation ($e^+e^-$ and pion production, (the GZK effect \citep{27, 28})).  It has also been suggested that the hadronic interaction of the CR with the atmosphere undergoes changes above certain energies which would affect the measured energy of the CR and hence produce an apparent change in the power law index \citep{Kaza03}.  A knowledge of CR composition as a function of energy would be invaluable in sorting out these effects.

High energy cosmic rays have long been known to be charged nuclei \citep{clay}, but determining the chemical composition at energies greater than 10$^{15}$ eV is especially difficult because of the low flux.  The High Resolution Fly's Eye cosmic ray observatory has a large enough aperture (3$\times$10$^{2}$ to 5$\times$10$^{3}$ km$^{2}$ str) but it observes the extensive air shower (EAS) produced by the particle rather than detecting the primary itself, and thus must use an indirect method to study the composition.

The distribution of positions of shower maxima (X$_{max}$) in the atmosphere has been shown to be sensitive to the composition of cosmic rays \citep{heit}.  It is well known that for any particular species of nucleus, the position of shower maximum will deepen with increasing energy as the logarithm of the energy.  The slope, $d(X_{max})/d(logE)$, is known as the elongation rate \citep{51}.

While the details are dependent on the hadronic model assumed, all modern hadronic models give approximately the same elongation rate (between 50 and 60 g cm$^{-2}$ per decade of energy, independent of particle species) and agree within about 25 g cm$^{-2}$ on the absolute position of the average shower X$_{max}$ at a given energy for a given species.  The sensitivity of the X$_{max}$ method to composition comes from the fact that the mean X$_{max}$ for iron and protons is different by about 75 g cm$^{-2}$, independent of hadronic model with protons 
producing deeper showers with larger fluctuations.  A change in the composition from heavy to light would then result in a larger elongation rate than 50-60 g cm$^{-2}$ per decade, and a change from light to heavy would lead to a lower and even negative elongation rate.

Previous experiments (stereo Fly's Eye \citep{38}, HiRes prototype-MIA \citep{proto}) have shown evidence for an elongation rate of 80-90 g cm$^{-2}$ g cm$^{-2}$ per decade in the energy rage from $10^{17}$ to $10^{18.5}$ eV.  No significant information from air-fluorescence experiments has been hitherto available on the behavior of the elongation rate near $10^{19}$ eV and above.

The general dependence of X$_{max}$ on energy can be seen in a simple branching model where $N_{max} \propto E_{\circ}$ and $X_{max} \propto \ln(E_{\circ})$ \citep{48, heit}. In this model, if the primary particle is a nucleus the shower is assumed to be a superposition of subshowers each initiated by one of the A independent nucleons.  The primary energy must be divided among the A constituents, so in this case $X_{max} \propto \ln(A/E_{\circ})$. A more complete discussion leads to Linsley's expression for the elongation rate, $\alpha$, \citep{53, 54}, which is

\begin{equation}
\alpha = (1 - B) K \lambda \left [1- \frac {d(log\langle A \rangle 
)}{d(logE_{\circ})} \right ]
\label{lins}
\end{equation}
where K is a constant, $\lambda$ is the collision length, and B expresses the dependence of $\alpha$ on the hadron-air nucleus interactions.  
It includes both the energy dependence of the cross-section (and thus $\lambda$) and the energy dependence of the multiplicity and inelasticity \citep{98a, 52, 55}.

The technique for extracting the cosmic ray composition used in this paper reduces to comparing the X$_{max}$ distribution of the data, after appropriate cuts that guarantee good resolution in this variable, to simulated data, generated with either a proton or iron parent particle. The simulated data is the result of a detailed detector Monte Carlo and includes all the reconstruction uncertainties.

\section{THE HiRes DETECTOR}
\label{detector}

The High Resolution Fly's Eye Cosmic Ray Observatory (HiRes) has two
sites on the U.S. Army's Dugway Proving Ground in the West Desert of
Utah, about 90 miles from Salt Lake City.  The first site, HiRes-1, is
on Little Granite Mountain, the site of the original Fly's Eye
detector.  HiRes-2 is 12.6 km to the southwest on Camel's Back Ridge.
Dugway was chosen for its clean atmosphere and low light pollution.
Each site is on a hill, above the bulk of most haze in the atmosphere.
The two sites gather data independently, and the data can be analyzed
from each site in monocular mode or together in stereo mode.  Results
from a monocular analysis of the spectrum have been published
\citep{mono1, mono2}.  The accurate determination of the shower geometry is an essential first step to determining the shower profile, which gives the energy of the primary UHECR and the X$_{max}$ of the shower.  This work will present results from stereo analysis, which has an obvious advantage in determining the shower geometry and hence better resolution in energy and X$_{max}$.

HiRes is the realization of an extension of the method pioneered by
the original Fly's Eye experiment \citep{flyseye}.  As a
UHECR-initiated Extensive Air Shower (EAS) propagates through the
atmosphere, the charged particles excite nitrogen molecules, which
fluoresce.  The fluorescence yield and its spectrum have both been
measured \citep{bunner, kakimoto}.  The fluorescence yield is
about five photons per particle per meter.  The photons, mostly of
wavelength 300-400 nm, propagate isotropically from the shower core,
with the number of photons coming from a slice of the shower proportional to the number of charged particles in that slice.  In addition to air fluorescence, \v{C}erenkov light is produced by shower particles. The \v{C}erenkov cone in air is $2.5^{\circ}$ and the resultant light angular distribution, reflecting the multiple scattering of electrons, is strongly beamed forward along the 
EAS axis. Some of this light, however, will scatter at larger angles into the detector and needs to be subtracted from the total signal.

Air fluorescence detectors gather the photons and focus them onto arrays of photo-multiplier tubes (PMTs).  Each PMT views a solid angle of the sky, and a composite
view is recovered when the images from the constituent mirrors are combined. 

The HiRes Prototype detector, results from which are also summarized in this paper, was located on Little Granite Mountain \citep{hrnim}.  The prototype detector viewed from $3^{\circ}$ to $70^{\circ}$ in elevation and had an azimuthal coverage that overlooks the Michigan Muon Array (MIA) 3.4 km away.  Coincident hybrid data were collected from 1993 to 1996.  Results on composition near $10^{17}$ eV were published in \citep{proto}.

In 1997, the aperture for events at the highest energies was optimized
by redeploying the telescopes in a ring with full $360^{\circ}$
azimuthal coverage and $3^{\circ}$ to $17^{\circ}$ elevational
coverage \citep{mono2}.  HiRes-1 began taking data in May of 1997,
and the ring was complete in March of 1998. 

HiRes-2 has two rings, giving complete azimuthal coverage with an
elevation coverage from $3^{\circ}$ to $31^{\circ}$ \citep{fadc_nim}.
This analysis uses data from November of 1999, which is when HiRes-2 was
completely operational, to September of 2001. 

\subsection{Survey and Calibration}
\label{cal}

The basic HiRes element is a telescope consisting of a mirror of area
5.2 m$^2$ with an associated cluster of 256 PMTs and data-acquisition
(DAQ) electronics.  The PMT cluster is placed at a distance from the
mirror which optimizes the spot size. \citep{simpson}.  Taking into account the cluster obscuration, the effective area of each mirror is
3.72 m$^2$ \citep{mono2}.  The pointing directions of the individual
telescopes were surveyed at installation and are checked periodically
by observing signals produced by stars \citep{bergstar, stefan}.

The primary tool for the calibration of PMT sensitivity is a xenon
flash lamp mounted in a portable housing that can be moved from
telescope to telescope.  It is placed in the center of the
mirror, illuminating the cluster directly.  The lamp's output
has been measured to be stable to within 1/3\% flash-to-flash and
within 2\% over the course of a night \citep{72, 73}. 

To monitor the PMT response on a nightly basis, each site has a
frequency-tripled YAG laser which delivers light at 355 nm to each
PMT cluster via quartz optical fibers.  One fiber goes to the center of
each mirror, and one to each side of each PMT cluster.  The
mirror-mounted fibers illuminate the cluster directly allowing
monitoring of tube response, while the cluster-mounted fibers
illuminate the mirrors so mirror reflectivity can be
tracked\citep{77, 78}.

\section{ATMOSPHERIC CORRECTIONS}
\label{netatm}

Photons from an EAS travel through the atmosphere to reach
the telescopes, and understanding light transmission through the
atmosphere is vital.  The atmosphere can be considered a mix of molecules and aerosols.  Rayleigh scattering from molecules is well understood, as is the
atmospheric density profile.  Scattering from aerosols varies with
the aerosol content of the air and must be measured.
We characterize the aerosol scattering by the horizontal aerosol extinction
length, L$^a$, the vertical aerosol optical depth, VAOD, and by the
angular dependence of the scattering cross section (the phase
function).  Lasers with steerable beams, located at both HiRes-1 and HiRes-2 sites,
are used to sweep the HiRes aperture and determine the horizontal
extinction length, VAOD and phase function \citep{80, 81}.  A database of the hourly parameterizations of the atmosphere is used in Monte Carlo and for the reconstruction of every event for which it is available \citep{74}.  For the final data set, about three-quarters of the events were reconstructed with atmospheric parameters from the hourly database and the remainder were reconstructed with the average values (see Section \ref{atcut}). 

The laser tracks also give an indication of clouds in the aperture.
The track of a laser hitting a cloud will mushroom out, giving a
shorter, wider signature.  

Other tools for understanding the clarity of the atmosphere include
operator observations, infrared cloud monitors, and xenon flashers
\citep{79}. One inclined and ten vertical xenon flashers located between the two HiRes sites are fired every ten minutes and give important qualitative information about the atmosphere between the sites.

\section{DATA ANALYSIS}

\subsection{Reconstruction}
\label{recon2}

The initial steps in the data processing chain are documented by \citet{74} and \citet{65}.  First, calibration information is applied to the raw data. The relative timing information from each mirror is then converted to an absolute time as determined by GPS \citep{67}.  The individual mirror triggers are matched to form multiple-mirror events, and the multiple-mirror events from each site 
are time-matched to build stereo events. To separate noise events from track-like events, a Rayleigh filter is employed \citep{101}.  The probability that an event was created by random noise is required to be less than 0.1\%.

\subsection{Geometry}
\label{geo}

\begin{figure}
\epsscale{0.85}
\plotone{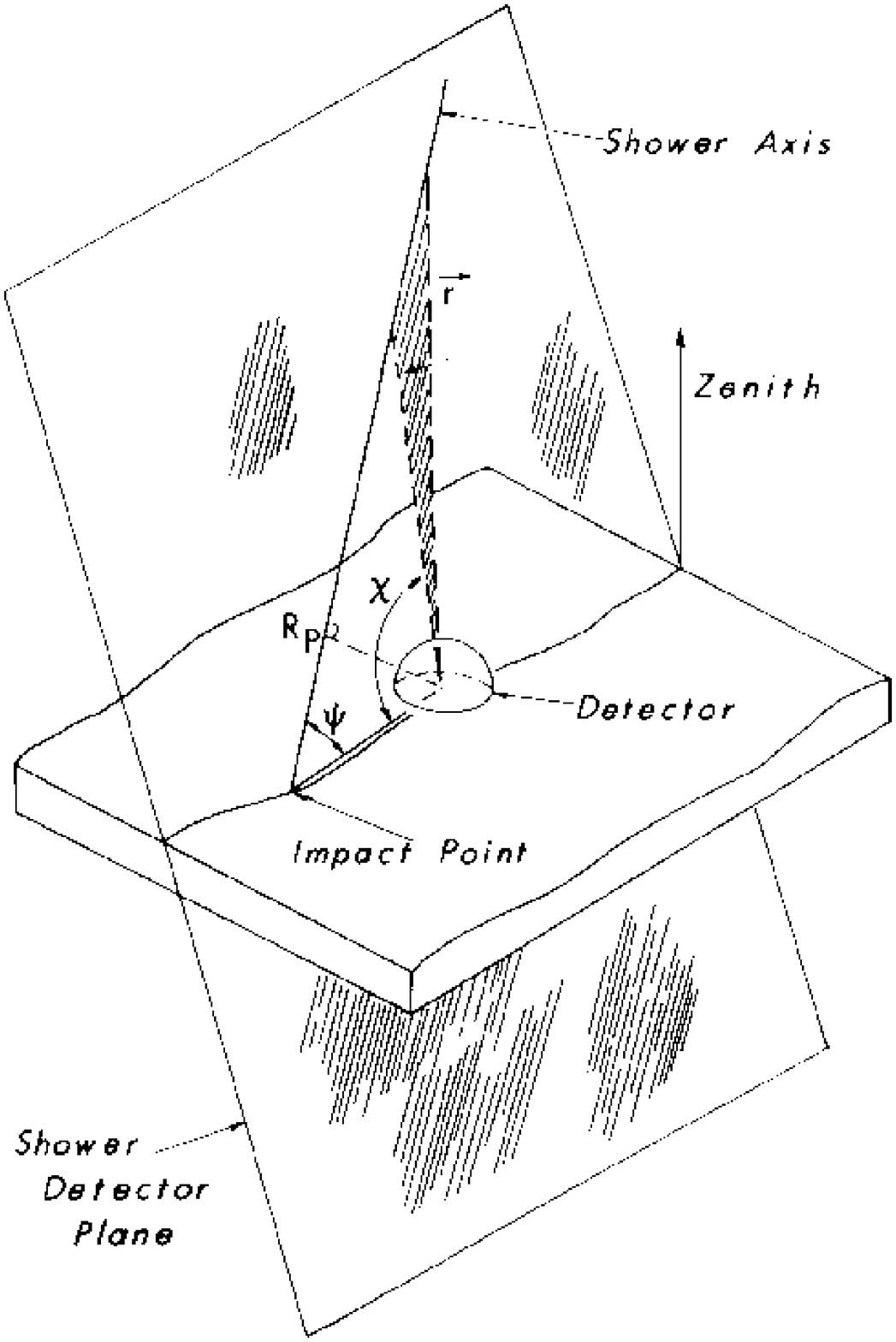}
\epsscale{1.0}
\caption{The Shower-Detector Plane.  The point of the detector and the line of the shower define a plane.\label{f1}}
\end{figure}

The axis of the EAS and the position of the detector uniquely define
the Shower-Detector Plane (SDP), as illustrated in Fig.
\ref{f1}.  The location and pointing direction of each PMT cluster have been measured \citep{bergstar, stefan}, so the pointing direction n$_i$ of each PMT is known and the SDP is easily found. 

Once the SDP is known for each site, the intersection of the planes
gives the direction and location of the EAS.  The next step in the reconstruction is to calculate the shower development profile. However, light arriving at the detector  is collected by discrete PMTs, each of which covers about  $1^{\circ}\times 1^{\circ}$ of the sky.  The signal from a longitudinal segment of the EAS is thus necessarily split among many PMTs.  For profile fitting, the signal must be re-combined into bins that correspond to the longitudinal segments of the EAS. The re-binned signal, corrected for atmospheric extinction and with \v{C}erenkov light subtracted is fit to a Gaisser-Hillas functional form 
(Eq. \ref{gh}). This form has been shown to be in good agreement with EAS simulations \citep{101, 89, 87} and with HiRes data \citep{102}.

\begin{equation}
N(X)=N_{max}{\left(\frac{X-X_{\circ}}{X_{max}-X_{\circ}}\right)}^
{(X_{max}-X_{\circ})/\lambda}\exp\left[\frac{(X_{max}-X)}{\lambda}\right].
\label{gh}
\end{equation}

\subsection{Angular Binning}
\label{ang}

A detailed description of the angular binning technique can be found
in \citet{74} and \citet{mythesis}.  With the incoming photon flux divided into the
angular bins, an ``inverse Monte Carlo'' method is employed to correct for acceptance.  A Monte Carlo shower with geometry corresponding to the event in question is generated with a Gaisser-Hillas profile (see Eqn. \ref{gh}).  Each photon, including both scintillation photons and \v{C}erenkov photons, is
individually traced up to the same point where the flux $\Phi$ was
computed from the data, giving $\Phi^{MC}$.  The atmospheric parameters described in Section \ref{netatm} are used to calculate attenuation and scattering.  If measured values of horizontal attenuation length and scale height are available in the data base for the hour during which the event occurred, those measured values are used.  Otherwise, the average atmospheric values are used \citep{83}.  N$_{max}$ and X$_{max}$ are then allowed to vary to minimize the $\chi^2$ with respect to the measured flux

\begin{equation}
\chi^2_{MC}=\sum_{i=1}^{j}{\frac{(\Phi_i-\Phi_i^{MC})^2}{\sigma_i^2}}
\label{imc}
\end{equation}
where j is the number of bins.  For this $\chi^2$, $\sigma$ is given by

\begin{equation}
\sigma^2_i=N^2_{pe_i}\sigma^2_{C_{A_{eff_i}}}+\sigma^2_{N_{pe_i}}
C_{A_{eff_i}}
\end{equation}
where N$_{pe}$ is the number of photoelectrons in the bin, C$_{A_{eff}}$ is a geometric correction factor for non-normal incidence, and 

\begin{equation}
\sigma^2_{N_{pe_i}}=N_{pe_i} + 40\space (pe/\mu sec)
\end{equation}
where 40\space (pe/$\mu$sec) is the average sky noise.

Eq.~(\ref{imc}) can be minimized for HiRes-1 and HiRes-2 individually,
or for both sites globally.  The N$_{max}$ and X$_{max}$ that minimize
Eq.~(\ref{imc}) (along with the parameters X$_0$ and $\lambda$) define
the shower profile.  The total number of charged particles is obtained
from the integral of Eq.~(\ref{gh}), and the energy is calculated by
multiplying the total number of charged particles by the energy deposited per charged particle.  A correction is made for unobserved energy averaging about 10\%.   

\subsection{Time Binning}
\label{time}

In addition to this analysis method, a reconstruction technique which takes advantage of the FADC timing at HiRes-2 has also been developed \citep{mono2}. It has the advantage of being less dependent on the details of phototube acceptance. We adapt it to generate one of the quality cuts described below.

In this method, the SDP are found as above and the EAS geometry obtained from the intersection of the SDP is assumed. A first estimate of the number of charged particles at the EAS in each HiRes-2 time bin is calculated assuming all of the photons reaching the detector are from air fluorescence. Note that each time bin contains the contributions of a number of PMTs determined by the effective optical spot size. The first guess of the number of charged particles at the shower obtained from the data is compared to the number of charged particles from a Gaisser-Hillas profile.  A scan through each X$_{max}$ and its associated most 
likely N$_{max}$ is performed to find the best fit.  \v{C}erenkov light is then introduced, based on the number of charged particles from the Gaisser-Hillas fit.  The \v{C}erenkov light is traced to the detector and subtracted from the signal, and the process is repeated to find a new X$_{max}$ and N$_{max}$.  The iteration continues until satisfactory agreement between the predicted data and real data is obtained.  The energy of the event is calculated from the shower profile exactly as in the angular binning technique.

\subsection{Stereo Reconstruction Cuts}
\label{stcut}

To insure that both detectors were working properly, only data files in which at least 20 xenon vertical flashers events (see Section 3) are seen by both eyes are used for this analysis. We chose events of energy greater than $10^{18}$ eV  because the stereo aperture is rapidly decreasing below this energy.  The statistical reach (defined as at least 4 events per bin) of this sample of data corresponds to a maximum bin at $10^{19.4}$ eV. Additional loose cuts described in Table \ref{cut_table} were used to remove obviously badly reconstructed events and stereo mismatches without biasing the data sample.  For the period between November of 1999 and September of 2001, 728 events met all of the above criteria and were subjected to the atmospheric cuts.

\subsection{Atmospheric Cuts}
\label{atcut}

For this analysis, any event with a corresponding VAOD measurement of larger than 0.1 was cut.  However, because of equipment down time, the atmosphere was not measured for every hour for which we have data. In that case the operators' log comments and measurements of laser track length vs. width (see Section \ref{netatm}) were searched.  Periods during which the operators' comments indicated bad weather and/or the length vs. width of the laser tracks indicated that the aperture was cloudy were discarded.  Of the 553 events comprising the final data set, 419 had atmospheric database entries which were used for reconstruction.  The remaining 134 events had no database entry but occurred during good weather and were reconstructed with parameters corresponding to average atmospheric conditions.

\subsection{Quality Cuts}
\label{qualcut}

All remaining events were manually scanned using an event display.  The cuts listed in Table \ref{cut_table} produce a data set that contains real events during periods of operation where the atmosphere was well understood and the detector was working well. We use the Monte Carlo simulated data to define these loose cuts to insure that sufficiently precise determination of event energy and X$_{max}$ is had. To establish these cuts, a Monte Carlo set of 8341 events was 
generated with a Fly's Eye Stereo spectrum for these resolution studies. The input composition for the Monte Carlo was nearly equal numbers of protons and iron nuclei, evenly divided between the QGSJet and SIBYLL hadronic interaction models to reflect the extremes of possible composition and hadronic interaction models.  Table \ref{cut_table} summarizes the resultant cuts.

Proper reconstruction of the shower profile for events pointed at the detector is especially dependent on the modeling of the forward-beamed \v{C}erenkov light and its atmospheric scattering. A minimum viewing angle cut is applied to minimize this problem. The cut on opening angle between the SDPs is necessary because the geometry obtained by the intersection of the SDPs is not well-constrained when the planes are nearly parallel. The Gaisser-Hillas profile can be fit to HiRes-1 or HiRes-2 data individually or to both globally. While the composition results reported here are from the global fit, the individual fits provide additional quality selection criteria.  Events for which the $\chi^2$ per degree of freedom for either of the individual fits is larger than 20 are cut.  For the global fit, the $\chi^2$ cut is 15.

Events for which the disparity between the individual HiRes-1 and HiRes-2 X$_{max}$ fits is more than 500 g cm$^{-2}$ are discarded.  Similarly, each HiRes-2 X$_{max}$ from the time binning technique described in Section \ref{time} is compared to the X$_{max}$ from the angular binning global fit, and events differing by more than 500 g cm$^{-2}$ are cut. Finally, the geometric uncertainty 
component of the X$_{max}$ error described in Section \ref{dmc} is required to be smaller than 400 g cm$^{-2}$ for the individual HiRes-1 and HiRes-2 fits and smaller than 200 g cm$^{-2}$ for the global fit. Confidence that the fit to Eq.~(\ref{gh}) found the correct X$_{max}$ is bolstered when one of the sites sees both the rise toward  and fall from X$_{max}$.  The bracketing cut ensures this by requiring that the measured X$_{max}$ is no more than 60 g cm$^{-2}$ beyond the visible track.

\begin{figure}
\plotone{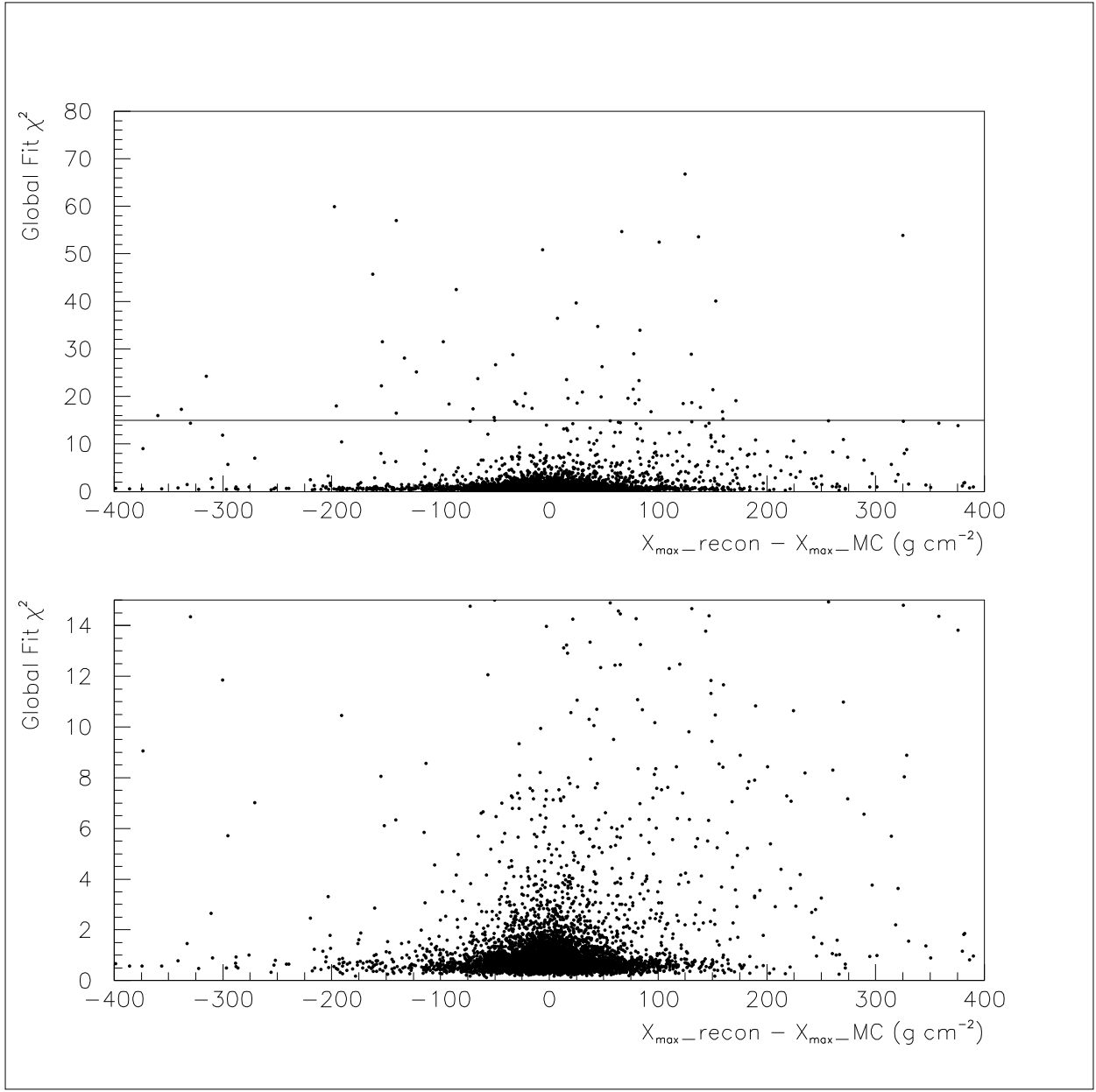}
\caption{Global Fit $\chi^2$ for reconstructed Monte Carlo events.  The horizontal line in the top plot delineates the cut at $\chi^2$ = 15.  The lower plot shows an expanded view of the region remaining after the cut.\label{f2}}
\end{figure}

\begin{figure}
\plotone{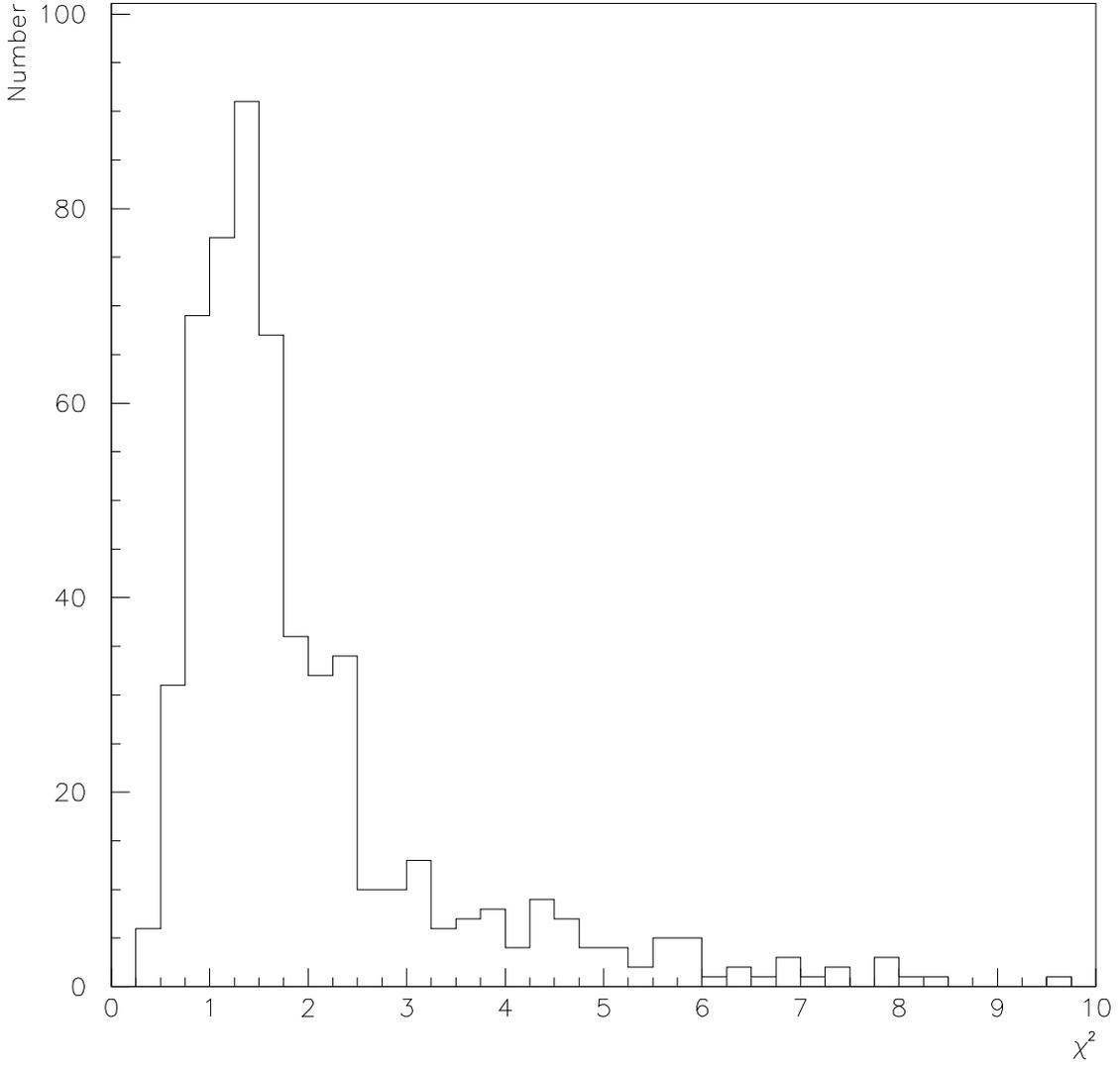}
\caption{Global Fit $\chi^2$ for the data.  After the application of all cuts, none of the data had a $\chi^2$ greater than 10.\label{f3}}
\end{figure}

As described below, in addition to the individual measurements of 
shower X$_{max}$ from HiRes-1 and HiRes-2 data, a global X$_{max}$ fit is performed utilizing all available data.  This global fit is what is finally used to determine composition.  The effect of these cuts on this variable is 
exemplified by Fig. \ref{f2} which shows the Monte Carlo global X$_{max}$ resolution as a function of $\chi^2$ and the cut location.  Fig. \ref{f3} shows the $\chi^2$ distribution for data after all the cuts are applied.

\begin{figure}
\plotone{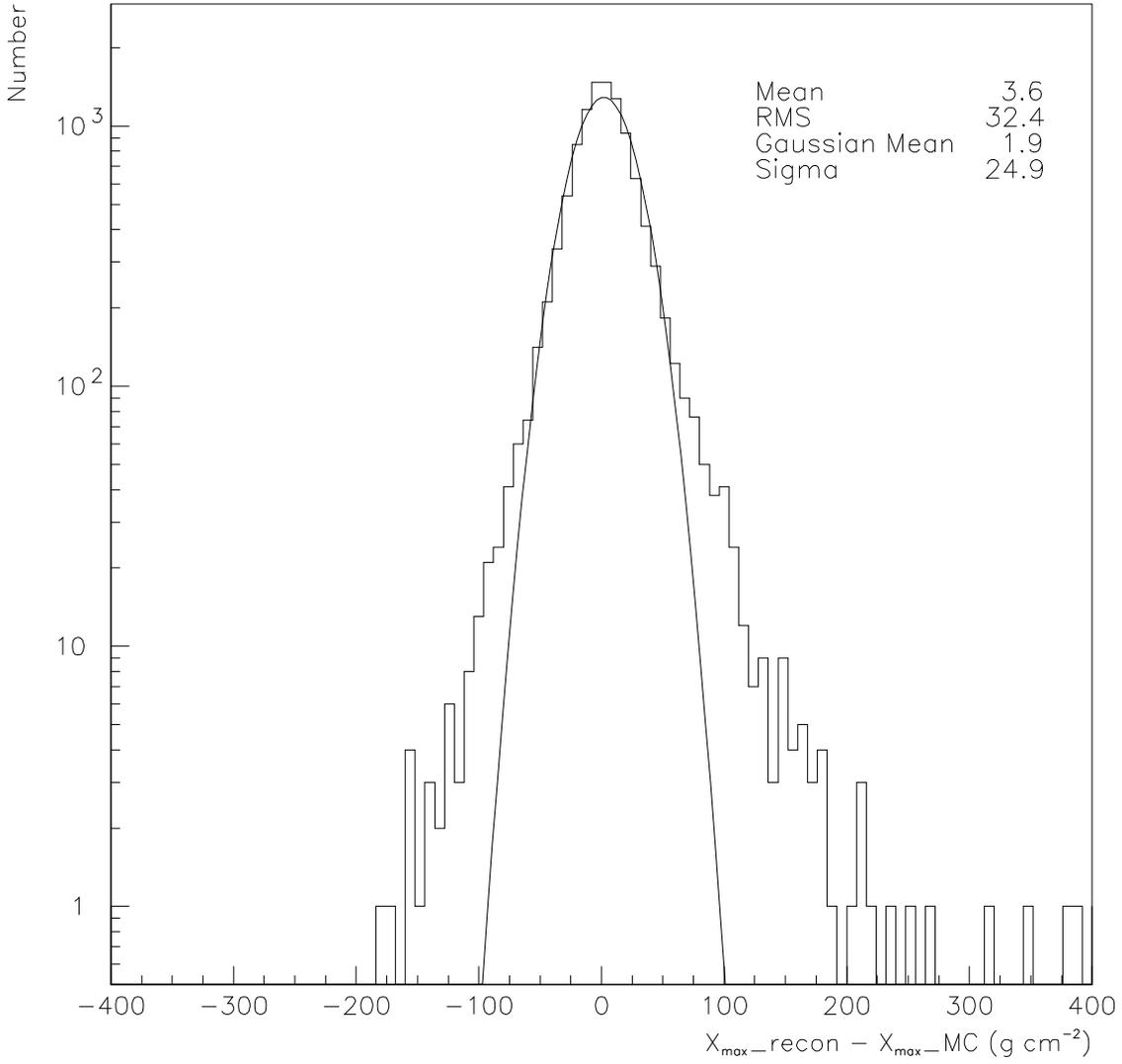}
\caption{X$_{max}$ resolution of the detector after all cuts, as determined by the detector simulation.  X$_{max}$\_MC is the thrown X$_{max}$ of the shower, and X$_{max}$\_recon is the reconstructed X$_{max}$.\label{f4}}
\end{figure}

\begin{figure}
\plotone{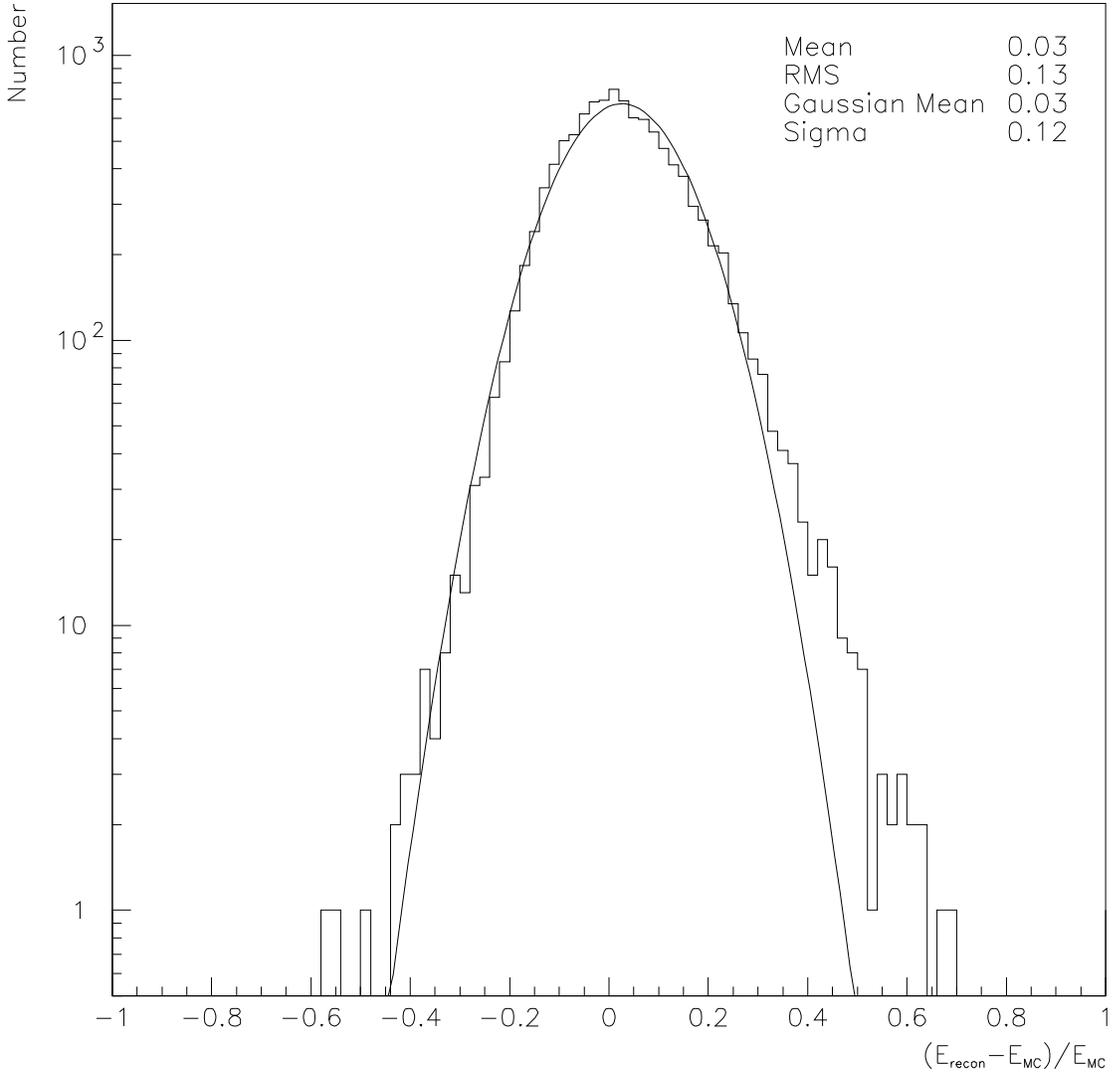}
\caption{Energy resolution of the detector after all cuts, as determined by the detector simulation.  E$_{MC}$ is the thrown energy of the shower, and E$_{recon}$ is the reconstructed energy.\label{f5}}
\end{figure}

\begin{figure}
\plotone{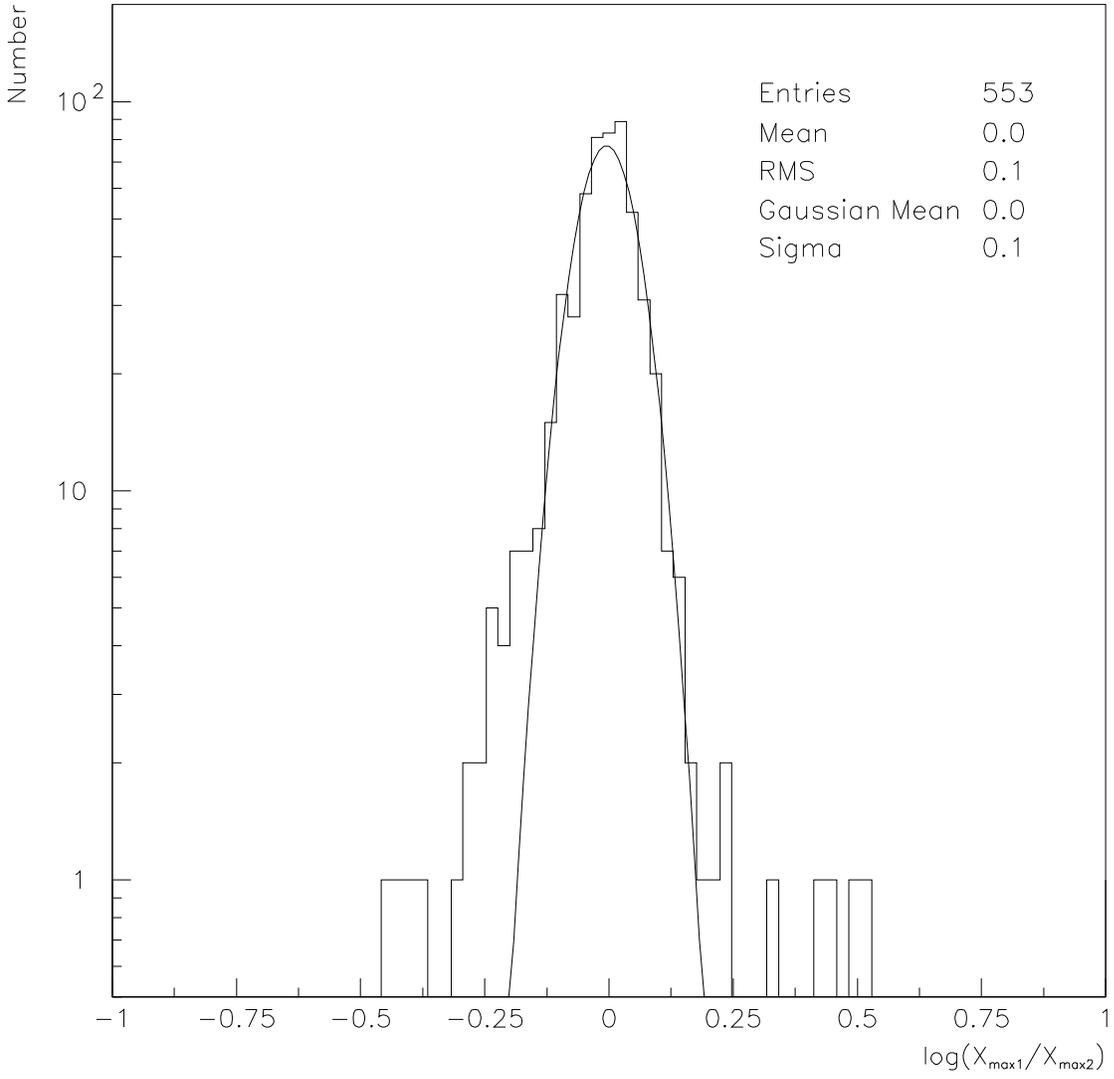}
\caption{Distribution of log(X$_{max1}$/ X$_{max2}$) in the data.  Compare Fig. \ref{f7}.\label{f6}}
\end{figure}

\begin{figure}
\plotone{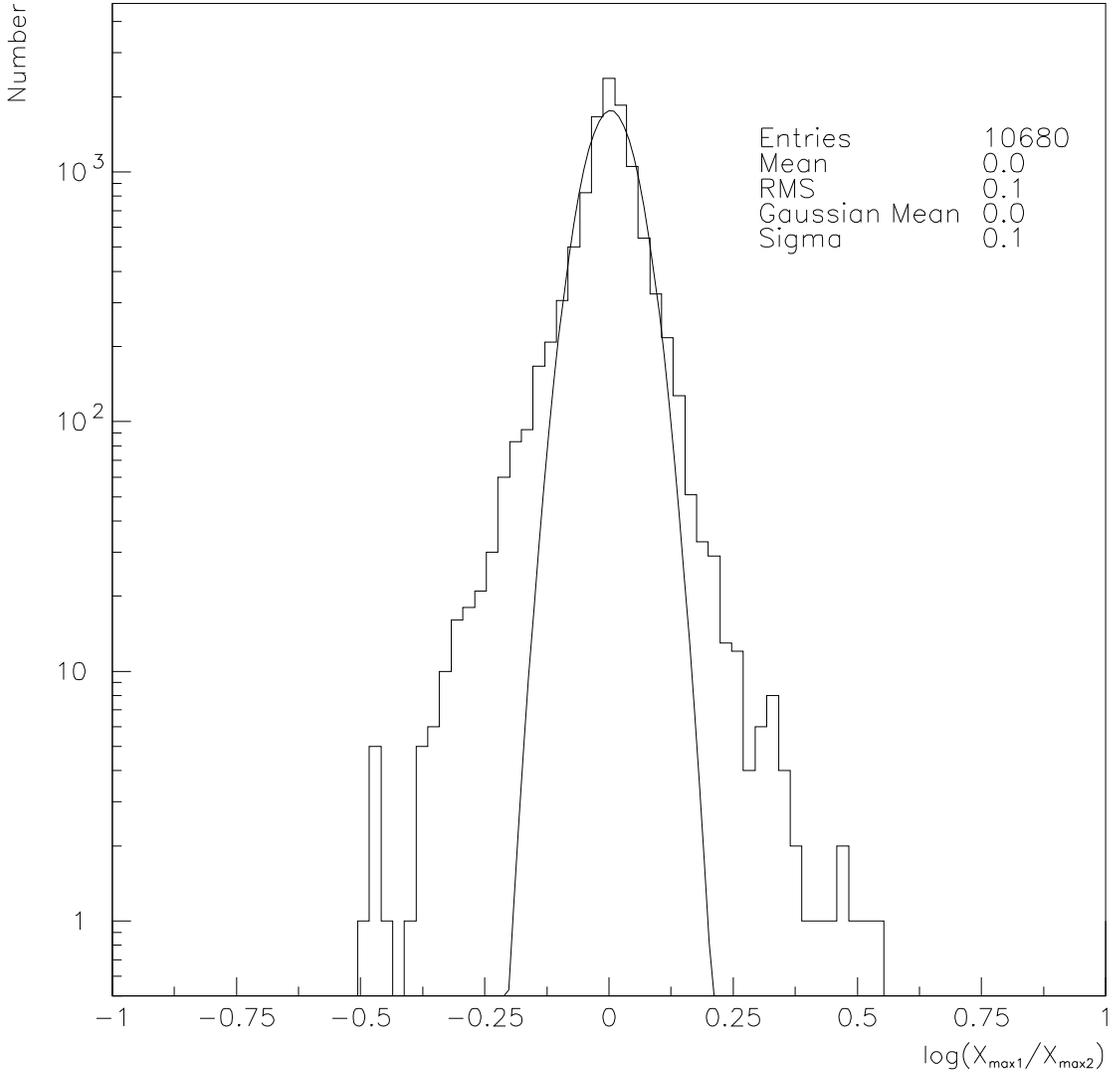}
\caption{Distribution of log(X$_{max1}$/ X$_{max2}$) for reconstructed Monte Carlo events.  Compare Fig. \ref{f6}.\label{f7}}
\end{figure}

\begin{figure}
\plottwo{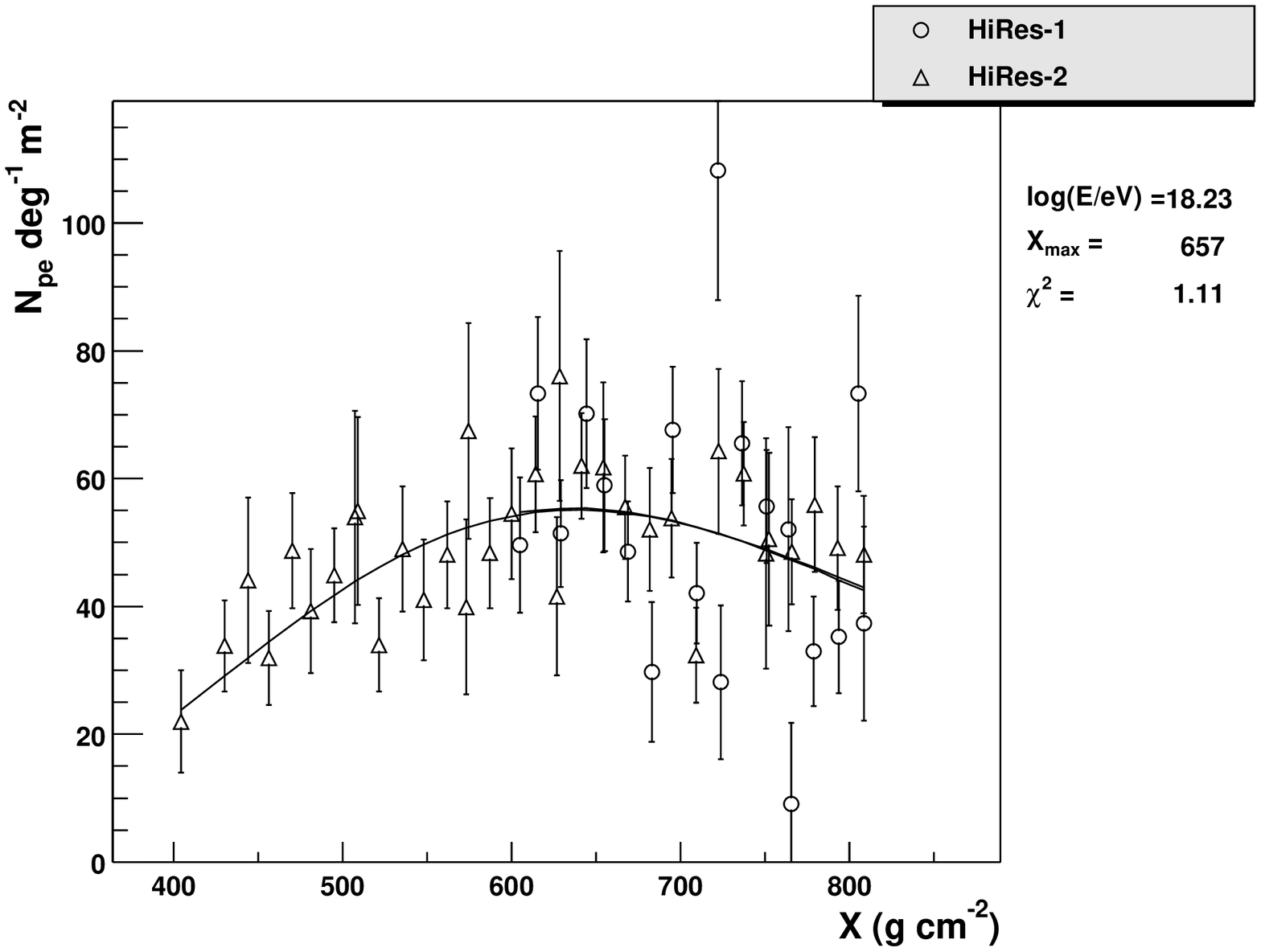}{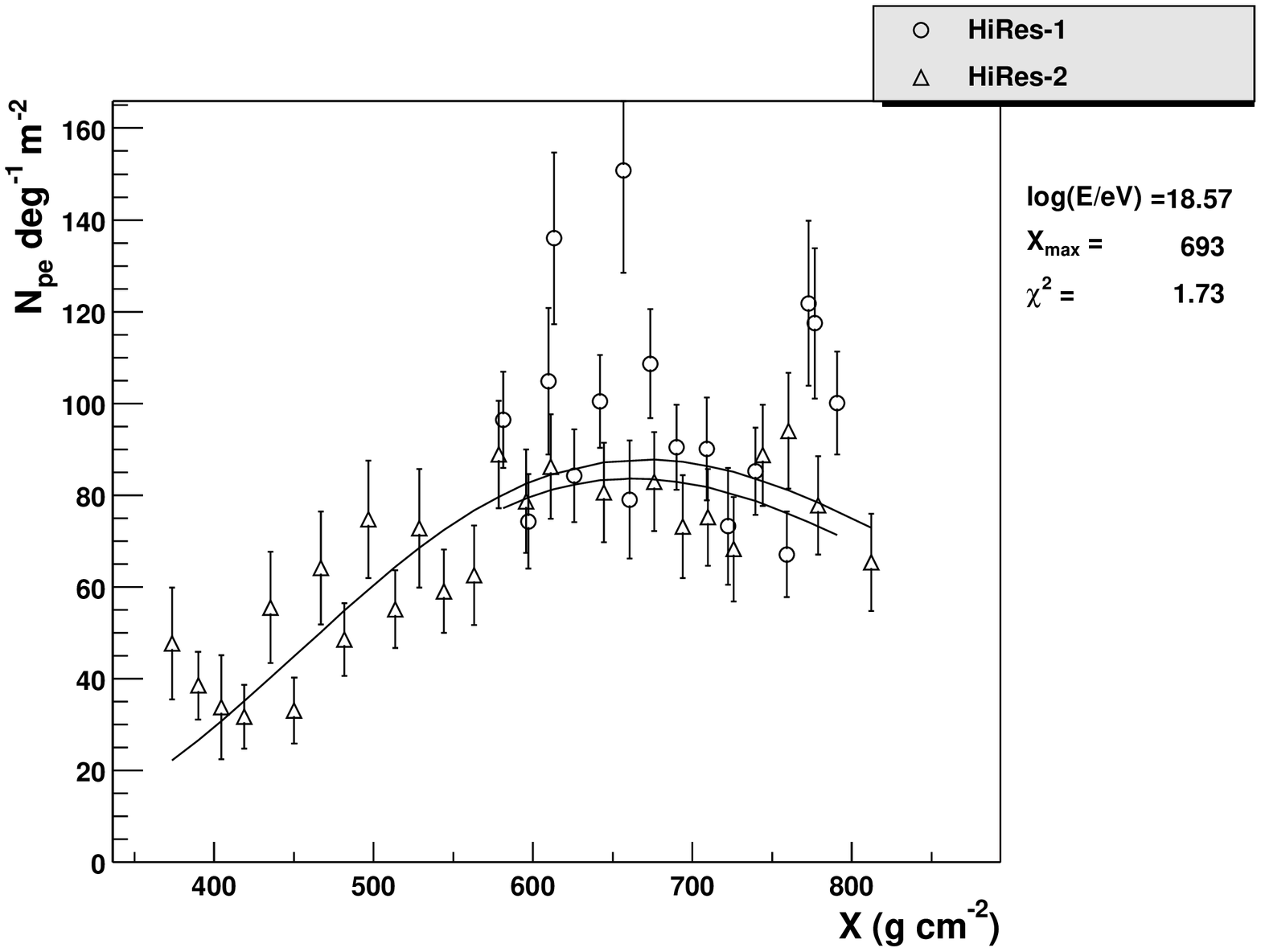}
\caption{Typical reconstructed shower profiles.  In the plots on the left, R$_{p1}$ and R$_{p2}$ were different by 0.5\%.  In the plot on the right, they differed by about 5\%.  The solid curve represents the global fit result.\label{f8}}
\end{figure}

\begin{figure}
\plottwo{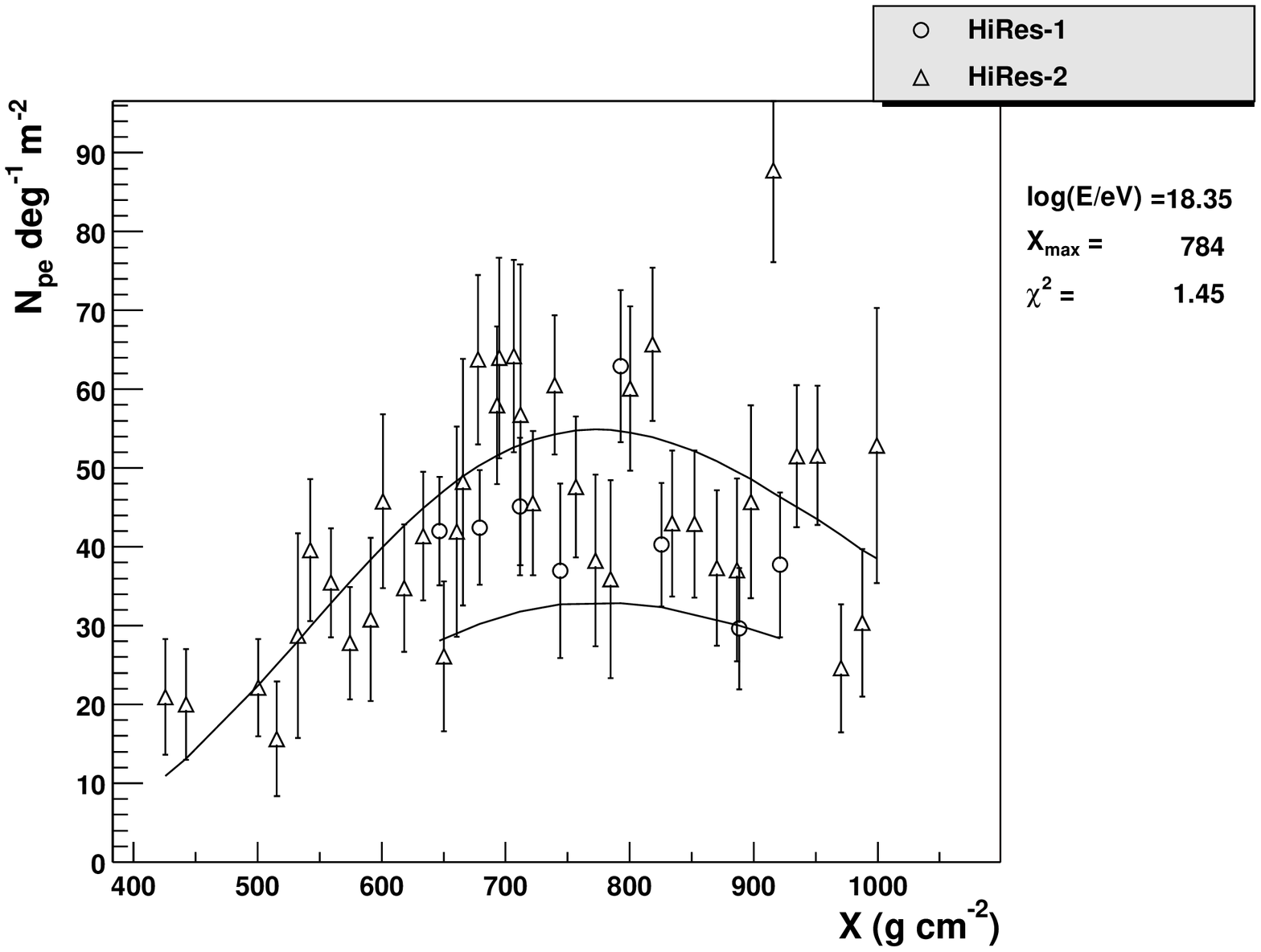}{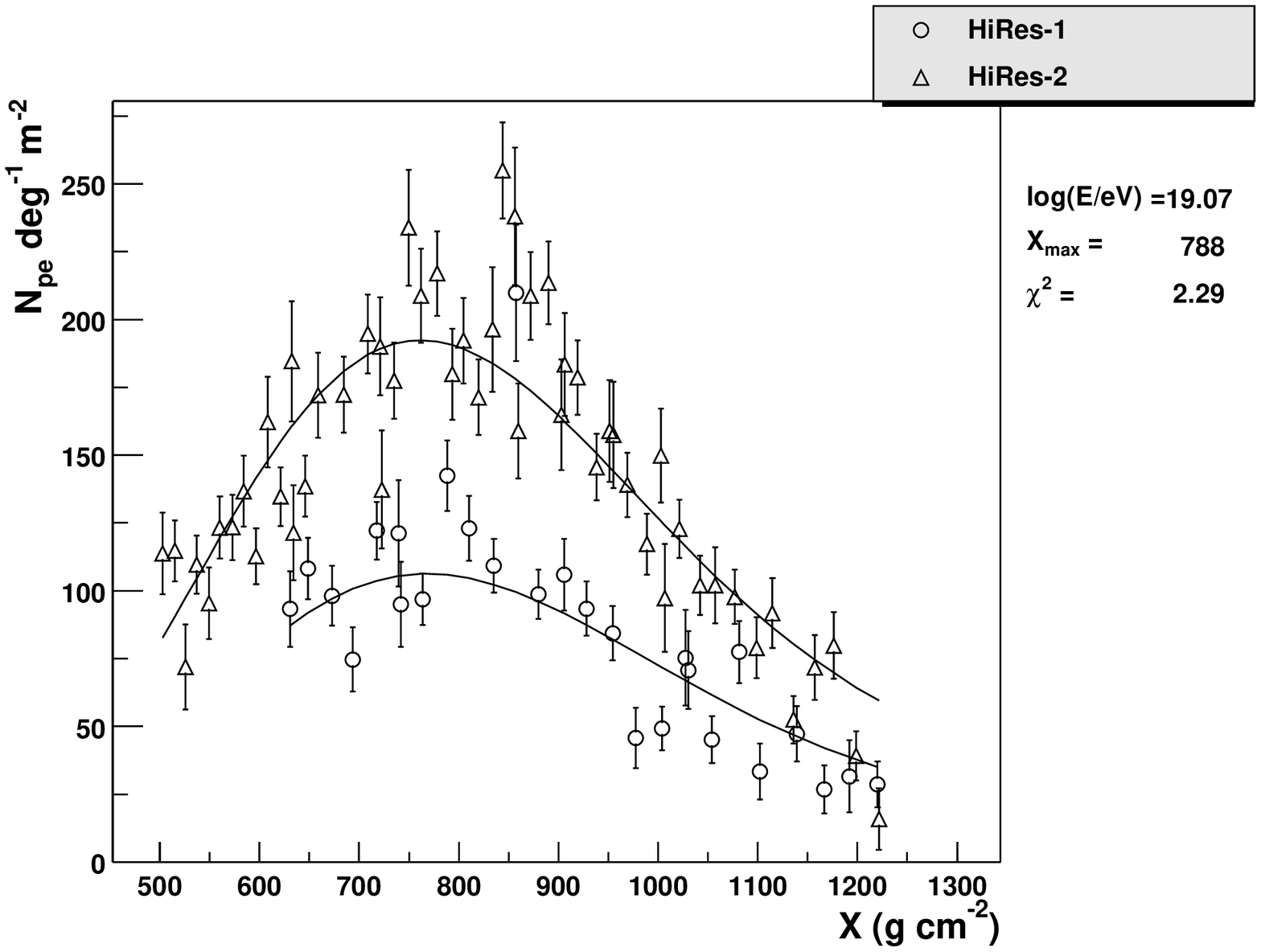}
\caption{Typical reconstructed shower profiles.  In both cases, the track was over 20\% further from HiRes-1 than HiRes-2.  The solid curve represents the global fit result.\label{f9}}
\end{figure}

\begin{figure}
\plottwo{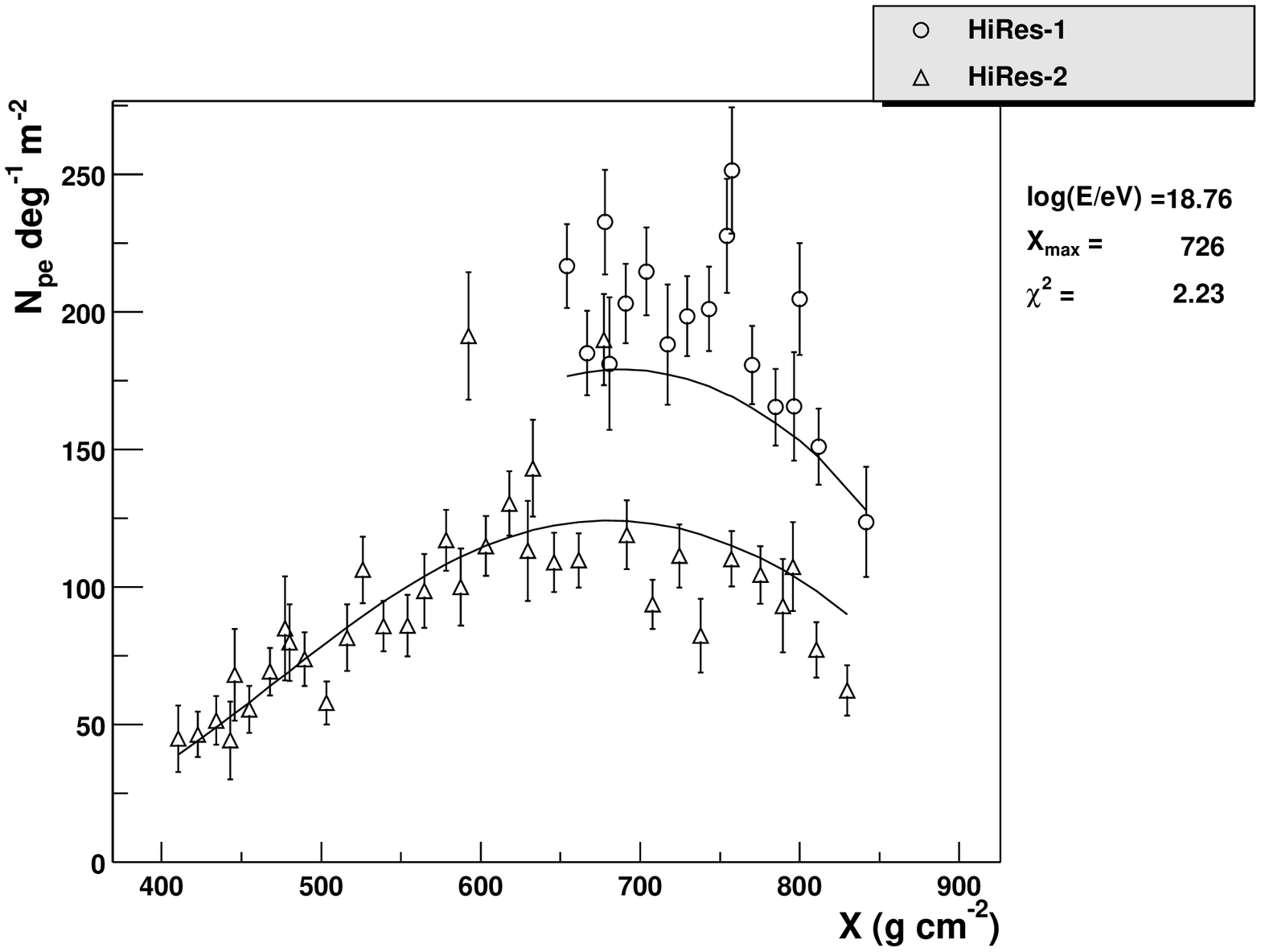}{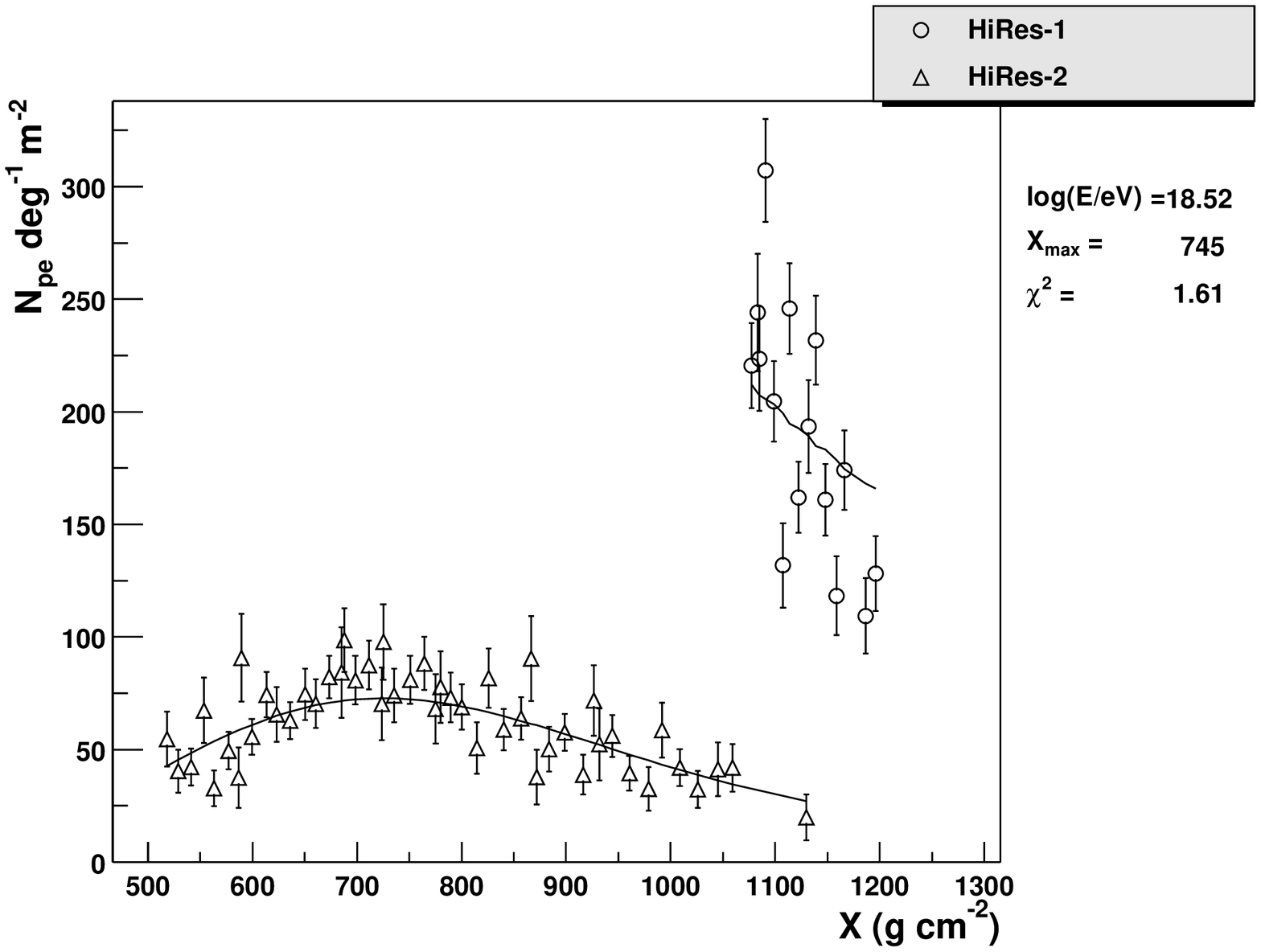}
\caption{Typical reconstructed shower profiles.  Even though HiRes-1 only observed a small portion of the shower, the SDP from HiRes-1 stringently constrains the global fit.  The energy balance depends in detail on atmospheric corrections.  In the example on the left, a perfect balance was not found, but the measured X$_{max}$ was unaffected.  The solid curve represents the global fit result.\label{f10}}
\end{figure}

553 events survive all of the above cuts.  The same cuts applied to the Monte Carlo set give a global X$_{max}$ resolution of 30 g cm$^{-2}$ and an energy resolution of 13\%. (See Figs. \ref{f4} and \ref{f5}). Fig. \ref{f6} shows the distribution of the ratio of log(X$_{max1}$/X$_{max2}$)  for the data.  The width of this distribution is a measure of the individual X$_{max}$ resolutions.  Because of shorter track lengths in the single ring, the HiRes-1 X$_{max}$ resolution is estimated to be 1.4 times worse than the HiRes-2 resolution.  Taking this into account, and propagating errors, the width of the distribution in Fig. \ref{f6} corresponds to a HiRes-2 resolution of about 50 g cm$^{-2}$, which is in accord with simulations.  Fig. \ref{f7} shows the same distribution for Monte Carlo events. The widths of the distributions are in good agreement. Figs. \ref{f8}--\ref{f10} show a representative selection of measured event profiles.

\section{EAS SIMULATION}
\label{corsi}

While cosmic ray hadronic composition presumably can range anywhere between the two extremes of pure proton and pure Fe, the 30 g cm$^{-2}$ resolution of the detector and the existence of significant shower fluctuations lead us to compare the data to a simplified two component model.  Events are generated using CORSIKA 6.005 and 6.010 \citep{89}, using both QGSJet01 \citep{87} and SIBYLL 2.1 \citep{86, 88} hadronic models for both protons and iron nuclei.

The differences in the hadronic interaction models are evident in the multiplicity, inelasticity, and hadron-air cross-section they predict.  Each of these directly affects the shower development.  The QGSJet multiplicity increases as $\log$(E), whereas the multiplicity in SIBYLL is below that of QGSJet at relevant energies and rises more slowly than $\log$(E) \citep{98}.

As a result of this multiplicity dependence, QGSJet showers would be expected to develop more quickly than SYBILL events.  Both models show an increase of inelasticity with energy \citep{87, 86, 88}, but QGSJet is more inelastic in the UHECR regime \citep{simpson}, again contributing to faster shower development\citep{98}. However, the hadron-air cross-sections in SIBYLL are larger than those of QGSJet at relevant energies.  The inelastic p-air cross-section in 
QGSJet rises approximately linearly with $\log$(E), while in SIBYLL it rises more rapidly \citep{98a}.  The nucleus-air cross-sections are comparable in magnitude for both models and rise slowly with energy \citep{98a, simpson}.

In all simulations, the CORSIKA EGS4 option was selected enabling
explicit treatment of each electromagnetic interaction for particles
above a threshold energy.  Electrons, positrons, and photons were
tracked down to energies of 100 keV.  Hadrons and muons were tracked
to 300 MeV.  All showers were initiated at $45^{\circ}$ to the
vertical, with sampling at 5 g cm$^{-2}$ of vertical atmospheric depth,
giving bins of about 7 g cm$^{-2}$ along the shower. 

Because of computational time requirements, simulated UHECR EAS must
be generated using a "thinning" approximation \citep{100}.  Numerous
studies have shown that setting the threshold for thinning at
10$^{-5}$ of the energy of the primary reduces computation time
without significantly affecting the results in the mean X$_{max}$ and
the elongation rate \citep{simpson, 101, 100, 62}.  The
thinning level for this work was set at 10$^{-5}$. 

Iron nucleus-initiated showers are expected to have smaller
shower-to-shower fluctuations than proton-initiated showers.  Studies
with shower generators, including CORSIKA, have shown that generating
as few as 200 iron showers at a given energy is sufficient to study
primary composition parameters, whereas 500 proton showers are needed
\citep{62}.  For this study, at least 400 iron showers and 500 proton
showers were generated using each hadronic interaction model in each
0.1 step of $\log$(E/eV) from E = $10^{17.5}$ to $10^{20}$ eV.

The use of thinning enhances fluctuations near shower maximum so that choosing the CORSIKA output bin with the largest number of charged particles will often yield a significantly incorrect X$_{max}$.  The input X$_{max}$ values used here were obtained using a weighted-average smoothing process.

For this work, a Monte Carlo library containing the complete profiles (number of charged particles as a function of atmosphere depth) for all 400+ iron showers and 500+ proton showers  generated at each energy was constructed.  Shower generation in the detector Monte Carlo is accomplished by sampling a shower profile in the library, making no a priori assumption about the shape of the shower profile.

\section{DETECTOR SIMULATION}
\label{recon}

A detailed Monte Carlo simulation of light production, atmospheric transmission, and detector response was developed in conjunction with the reconstruction routines, allowing the study of the reconstruction code's ability to correctly recover information about the primary particle.

\subsection{Monte Carlo Program}
\label{mc}

The full Monte Carlo simulation of a cosmic ray event begins with the generation of an EAS.  The longitudinal development of each CORSIKA shower is stored in the library described above.  After selecting a primary particle energy from the input spectrum, (based on the Fly's Eye Stereo Spectrum \citep{flyseye}), the Monte Carlo randomly selects a shower from the library energy bin closest to the desired energy and interpolates.  The zenith angle and distance from the detector are 
then chosen from random distributions.  In each atmospheric depth bin, scintillation light and \v{C}erenkov light are calculated based on the number of charged particles in that bin.  The propagation of the light to the detector is then simulated accounting for molecular and aerosol scattering.

The photon flux reaching the detector is distributed among the PMTs by a detailed ray-tracing program which has been checked by examining the point spread function of stars in the night sky \citep{bergstar, stefan}.  The ray tracing accounts for PMT cluster obscuration of the mirror, mirror shape, mirror reflectivity, and UV filter transmission.  Cracks between PMTs are also simulated.  A PMT quantum efficiency curve (28\% at 355 nm), based on specifications provided by the manufacturers, is used to obtain the number of photo-electrons. A simulated signal is then generated using the gain of the PMT and its pre-amplifier. The triggering conditions are applied, with the resulting output stored in the same format as real data.  The full timing of each Monte Carlo photon, from production at the shower to the PMT face, is stored so that the trigger timing in the output is accurate. The Monte Carlo also simulates noise.  Sky noise and electronics noise are added to each signal following a Poisson distribution, and random noise tubes are added to each event with the same mean and sigma as in the actual data. The Monte Carlo events are then processed by the same stereo and time-binned reconstruction routines used for the data.

\subsection{Atmospheric Comparisons}
\label{atm_comp}

\begin{figure}
\plotone{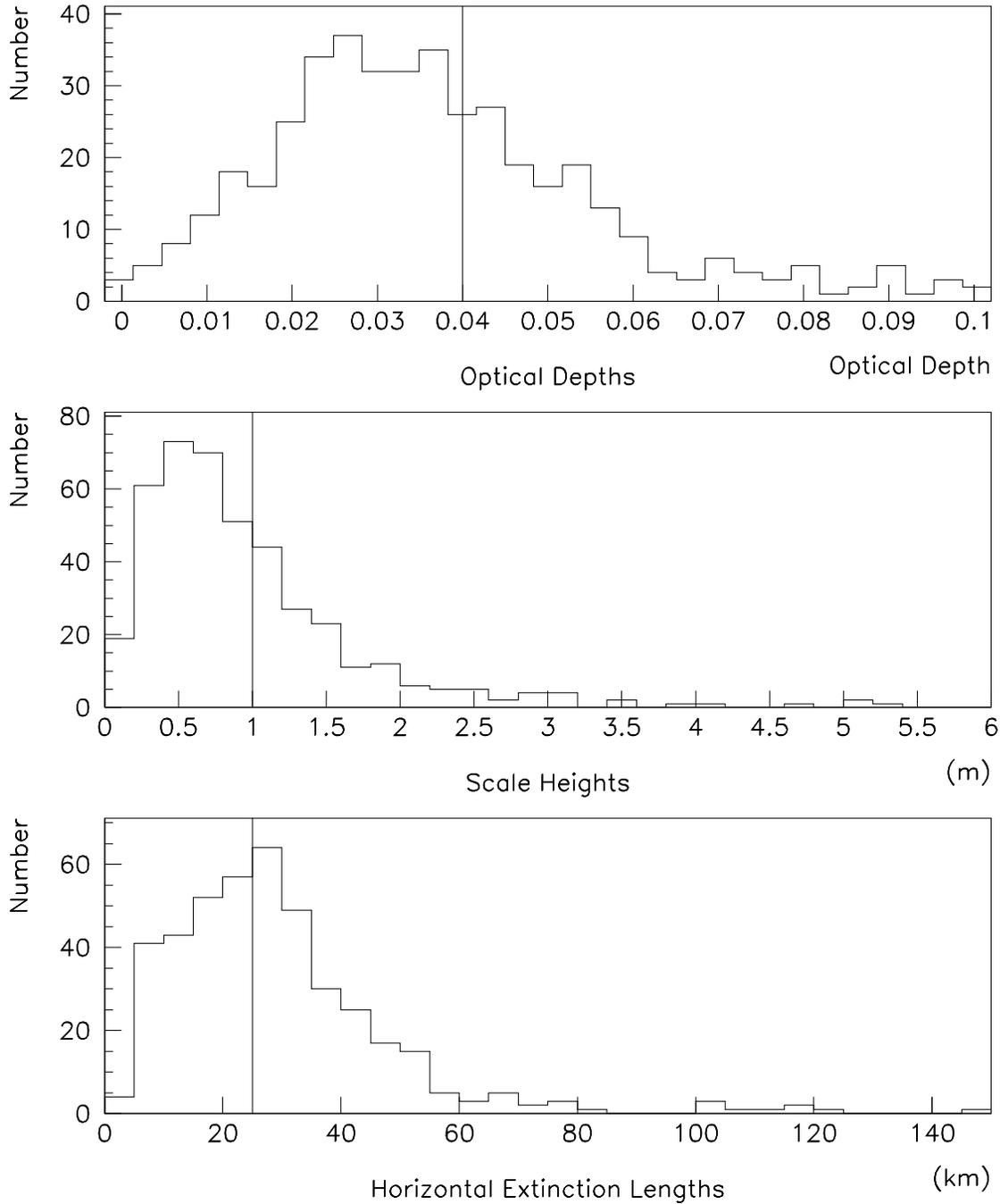}
\caption{Distributions of atmospheric parameters.  The vertical lines show the quoted average values, which are pulled to the right by measurements which give optical depths greater than 0.1 and are therefore not shown.  If no aerosols were present, the Horizontal Extinction Length would be inifinte \citep{83}.\label{f11}}
\end{figure}

Real UHECR events occur in whatever atmospheric conditions present themselves in the aperture. Hourly atmospheric parameters are available for most of the data.   Fig. \ref{f11} shows the distribution of atmospheric parameters in the database for VAOD of less than or equal to 0.1 (corresponding to the data cut). We use this distribution for thrown events.  The mean scale height is inferred from the mean horizontal attenuation length, the mean optical depth, and \citep{83}

\begin{equation}
AOD=\frac{H_s}{L^a}.
\label{saod}
\end{equation}

If no measurement exists in the database, events are reconstructed with the average atmospheric description as discussed above. To simulate the error introduced by this, Monte Carlo events were generated with input atmospheric parameters sampled from the database and then reconstructed with the average parameters. Monte Carlo events were also generated with the average atmospheric 
parameters.  Resolutions after quality cuts using the two sets were indistinguishable.

\subsection{Data-Monte Carlo Comparisons}
\label{dmc}

If the Monte Carlo accurately models the detector, then the application of an event selection criterion will have the same effect on Monte Carlo events and data.  The Monte Carlo can then be used to determine resolution by reconstructing Monte Carlo events and comparing the results to the input parameters.  Additionally, the effects of selection cuts on the resolution can be studied.  To 
determine how well the Monte Carlo models the detector, comparisons were made between over 20,000 Monte Carlo events and 926 UHECR events.

\begin{figure}
\plotone{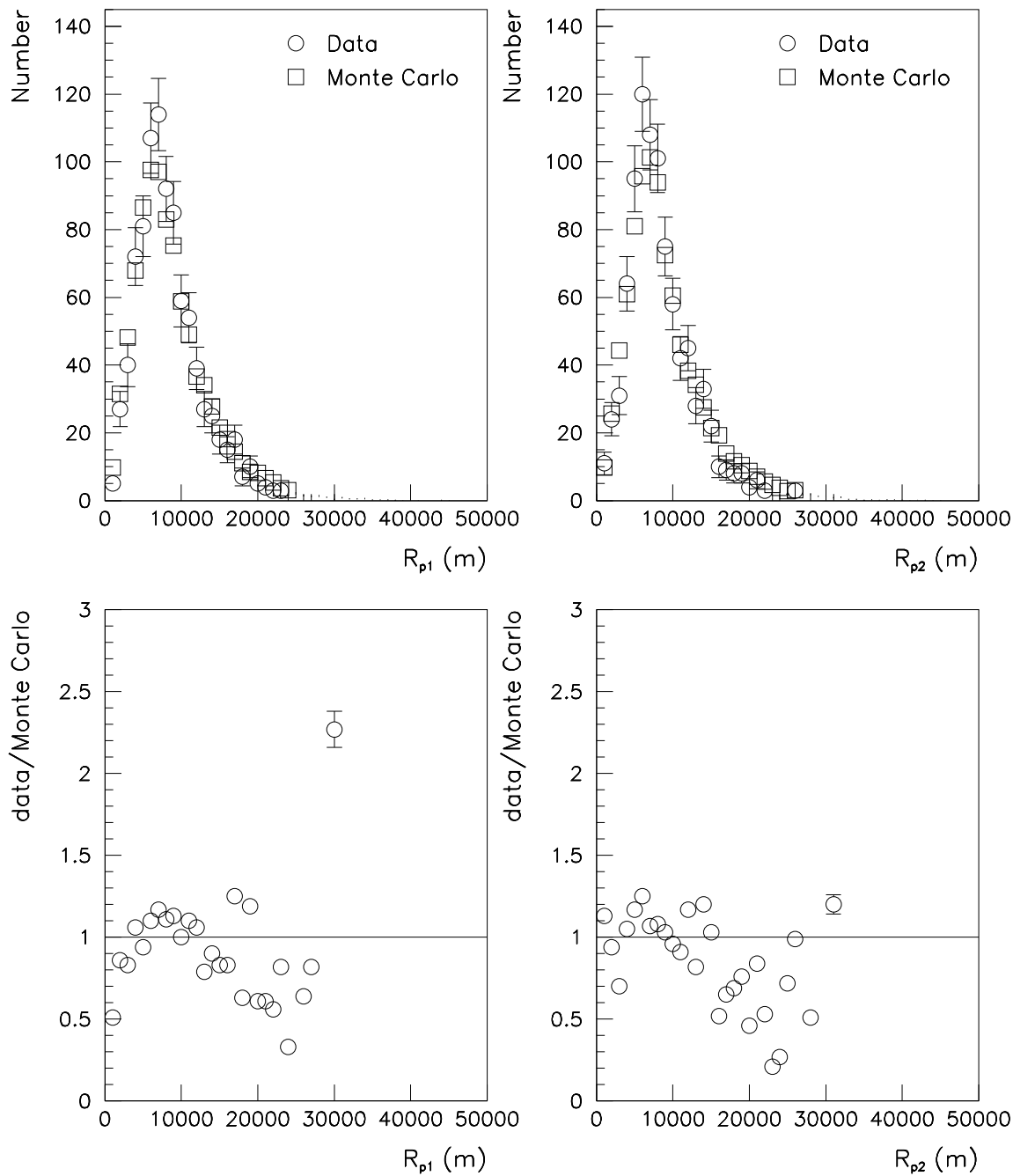}
\caption{Comparison of reconstructed R$_p$ in the data and the Monte Carlo. The top plots show the distributions in the data and the Monte Carlo, and the bottom plots show the ratio of data/Monte Carlo in each bin.  The similarities of the distributions indicate that the thrown events and the Monte Carlo are excellent representations of the data and the detector, respectively.\label{f12}}
\end{figure}

With the cuts described at the beginning of Section \ref{stcut} applied, Fig. \ref{f12} shows the data-Monte Carlo comparisons for distribution in R$_p$. The Monte Carlo histogram in the figure represents the results of reconstructing all 20,000+ Monte Carlo events, normalized to have the same area under the curve as the data histogram.  The bin-by-bin ratios shown are ratios of the data bin height to the normalized Monte Carlo bin height. Fig. \ref{f12} is typical of data-
Monte Carlo comparisons in other variables such as energy, zenith angle, $\psi$, tracklength in degrees and g cm$^{-2}$, and maximum single N$_{pe}$ deg$^{-1}$ m$^{-2}$ in each event. Good agreement between data and Monte Carlo is found in all cases.

\begin{figure}
\plotone{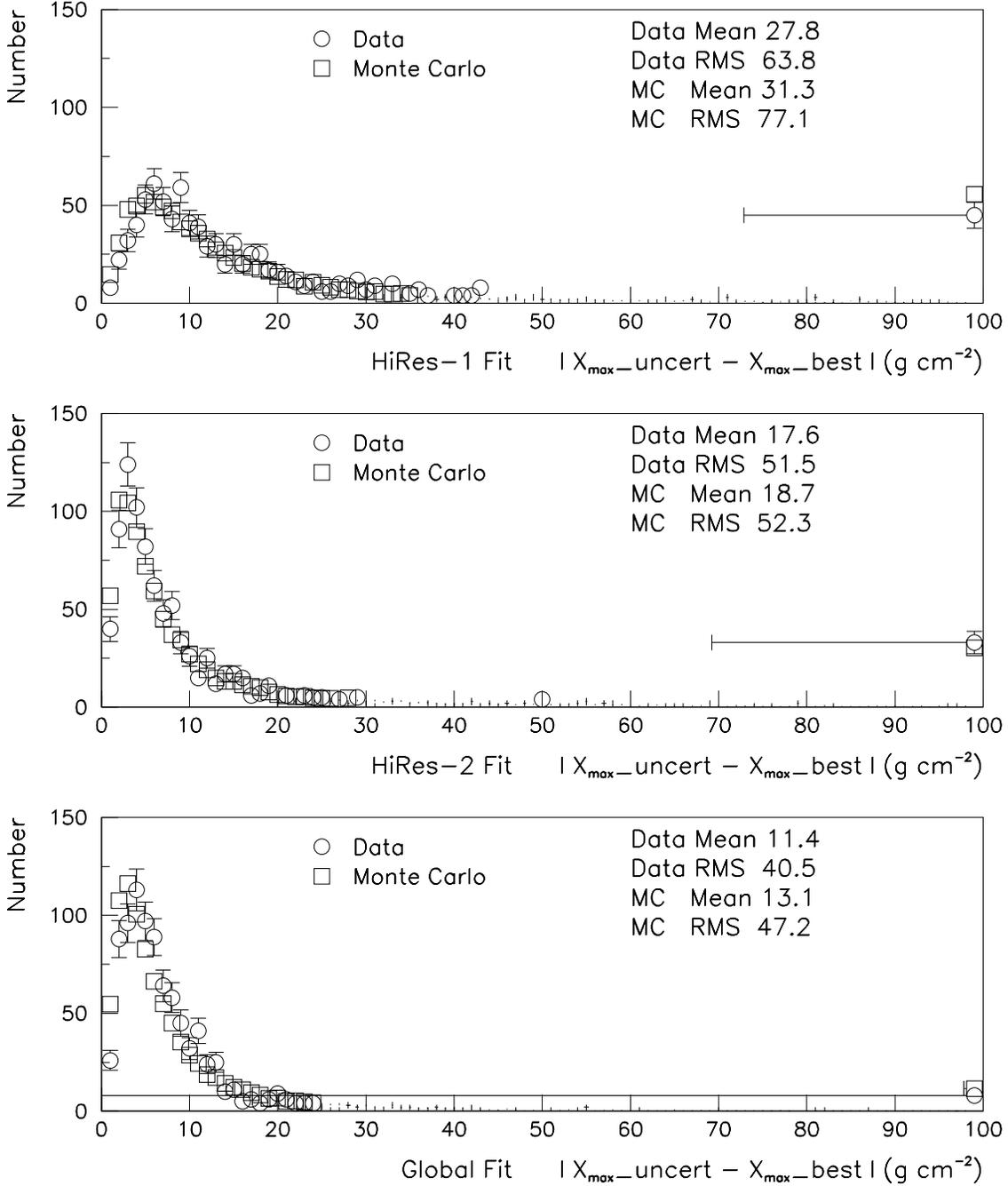}
\caption{Magnitude of X$_{max}$ error introduced by uncertainty in the SDP, for both the data and MC events, where X$_{max}$\_uncert is the X$_{max}$ found after the SDP is shifted by the worst-case uncertainty and X$_{max}$\_best is the X$_{max}$ using the best SDP.  The far right bin in each plot is an overflow bin.\label{f13}}
\end{figure}

An accurate profile determination depends on an accurate geometry. The SDP-finding routine returns the normal to the SDP as well as the uncertainties in each component of the normal.  For each event, the worst case error in each SDP was propagated through the reconstruction to give the uncertainty in X$_{max}$ due to the uncertainty in geometry.  Fig. \ref{f13} shows the distribution of these uncertainties for X$_{max}$ in the data and the Monte Carlo, again showing 
good agreement.

\begin{figure}
\plotone{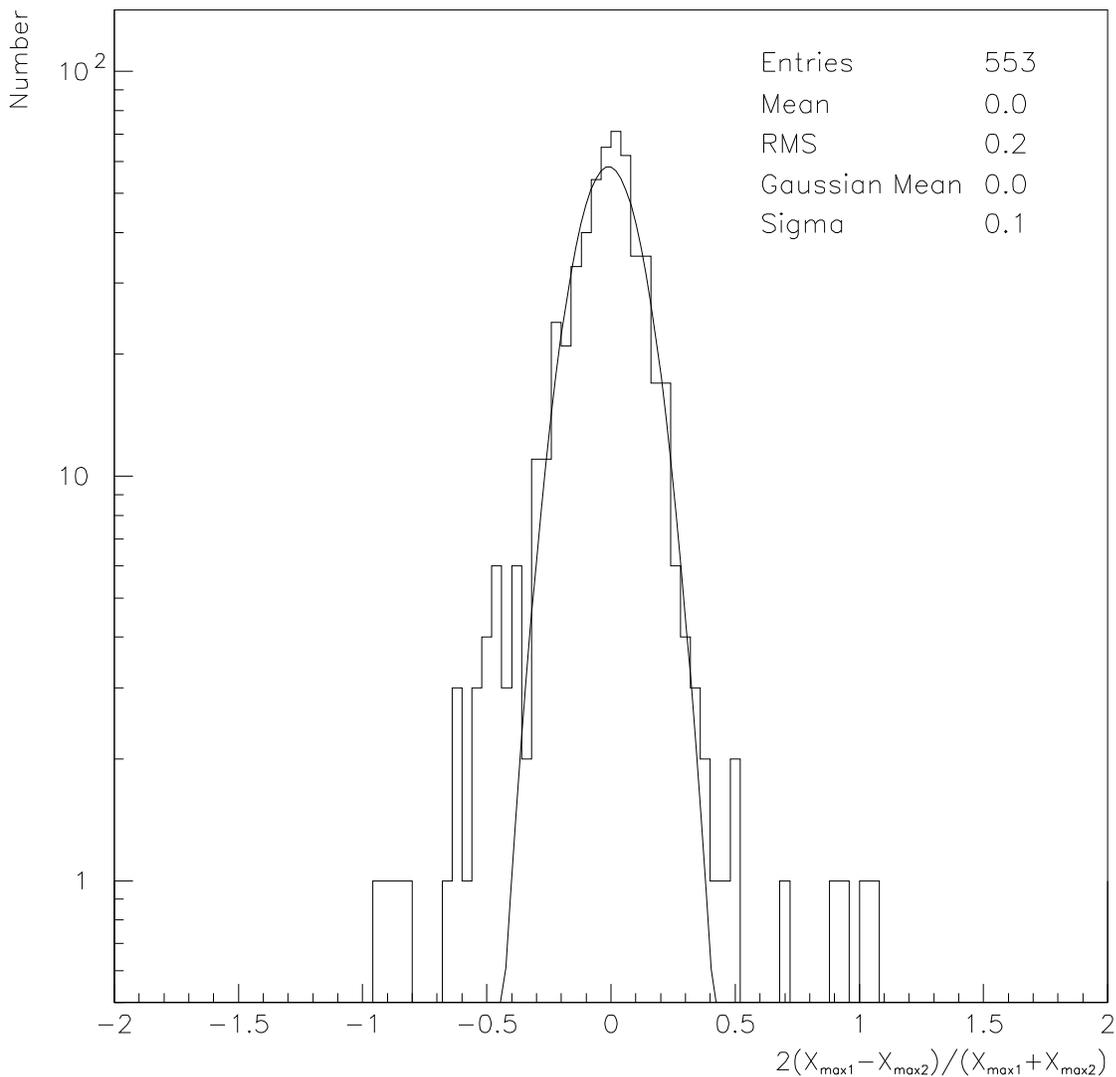}
\caption{X$_{max}$ Pull in the data.  The pull is defined as 2(X$_{max1}$ - X$_{max2}$)/(X$_{max1}$ + X$_{max2}$), where X$_{max1}$ and X$_{max2}$ are from the individual fits by HiRes-1 and HiRes-2, respectively.  Comparison with Fig.  \ref{f15} shows excellent agreement between the data and MC pulls.  Both the pull and log(X$_{max1}$ / X$_{max2}$) are differences between X$_{max1}$ and X$_{max2}$ and are statistically equivalent variables, so that Fig. \ref{f6} is identical to this plot except for a multiplier on the x-axis.\label{f14}}
\end{figure}

\begin{figure}
\plotone{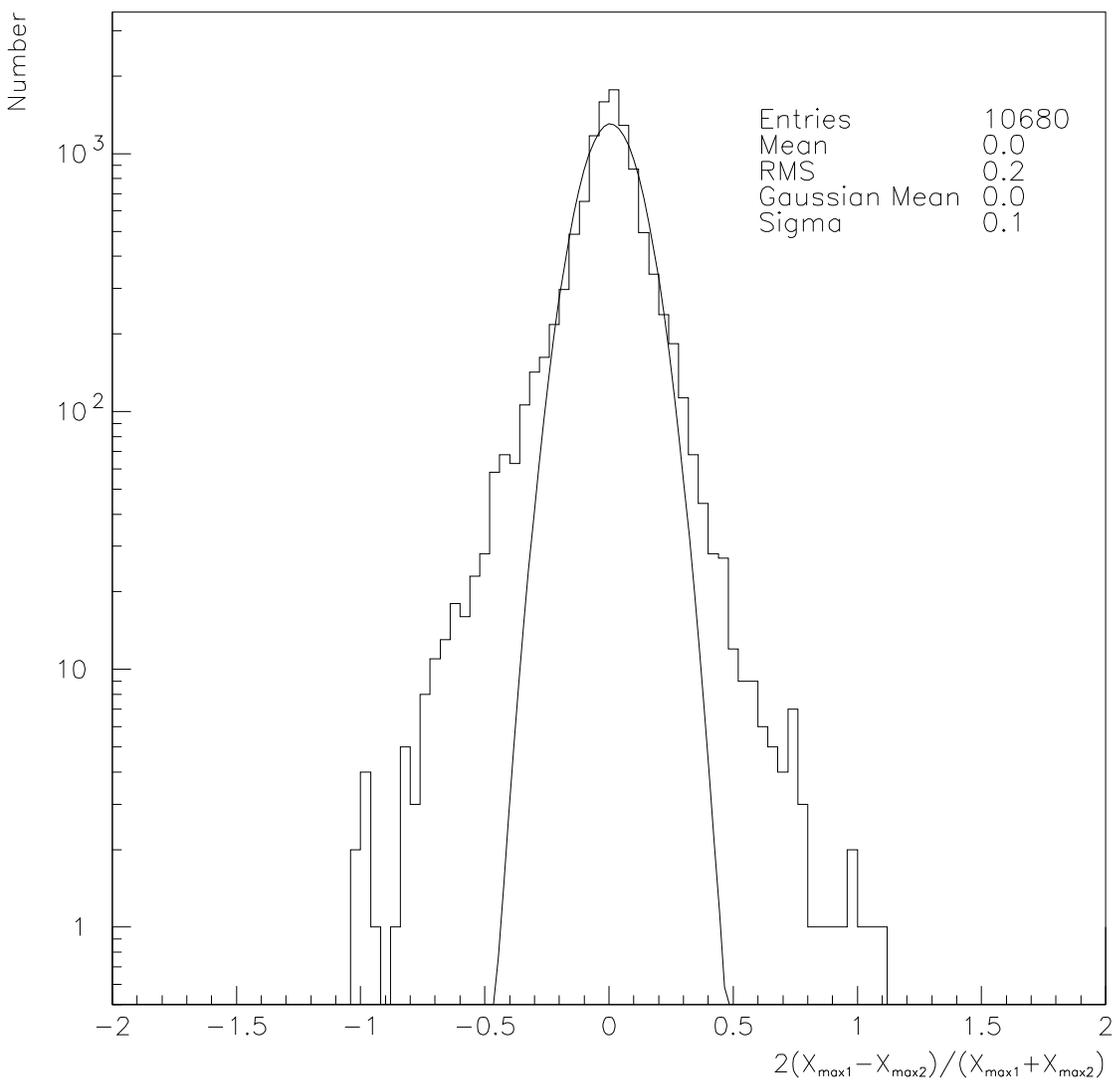}
\caption{X$_{max}$ Pull for reconstructed Monte Carlo events.  The pull is defined as 2(X$_{max1}$ - X$_{max2}$)/(X$_{max1}$ + X$_{max2}$), where X$_{max1}$ and X$_{max2}$ are from the individual fits by HiRes-1 and HiRes-2, respectively. Comparison with Fig. \ref{f14} shows excellent agreement between the data and MC pulls.  Both the pull and log(X$_{max1}$ / X$_{max2}$) are differences between X$_{max1}$ and X$_{max2}$ and are statistically equivalent variables, so that Fig. \ref{f7} is identical to this plot except for a multiplier on the x-axis. \label{f15}}
\end{figure}

\begin{figure}
\plotone{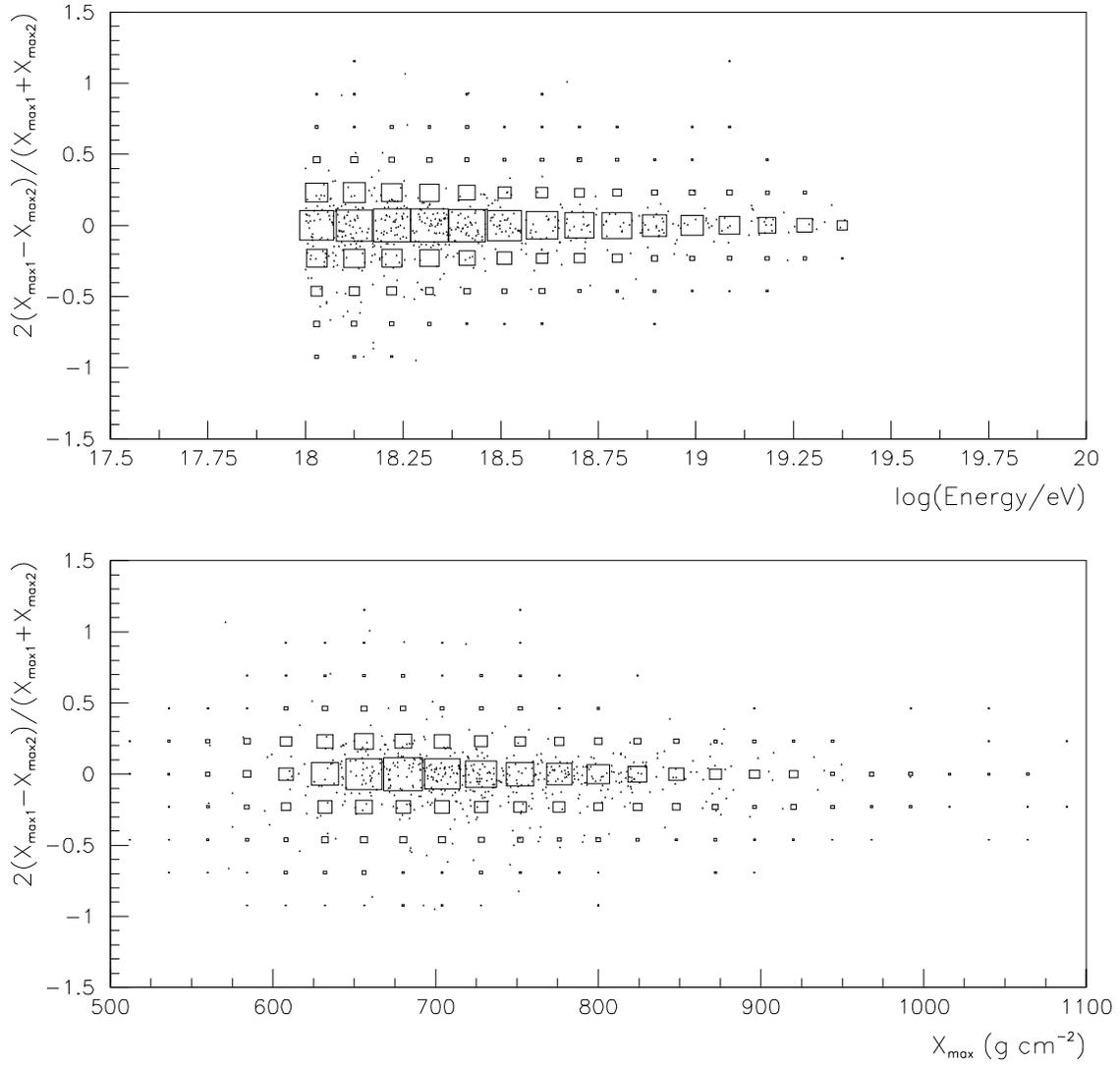}
\caption{Pull as a function of Energy and X$_{max}$.  The large boxes represent the Monte Carlo events, and the small triangles represent the data.\label{f16}}
\end{figure}

\begin{figure}
\plotone{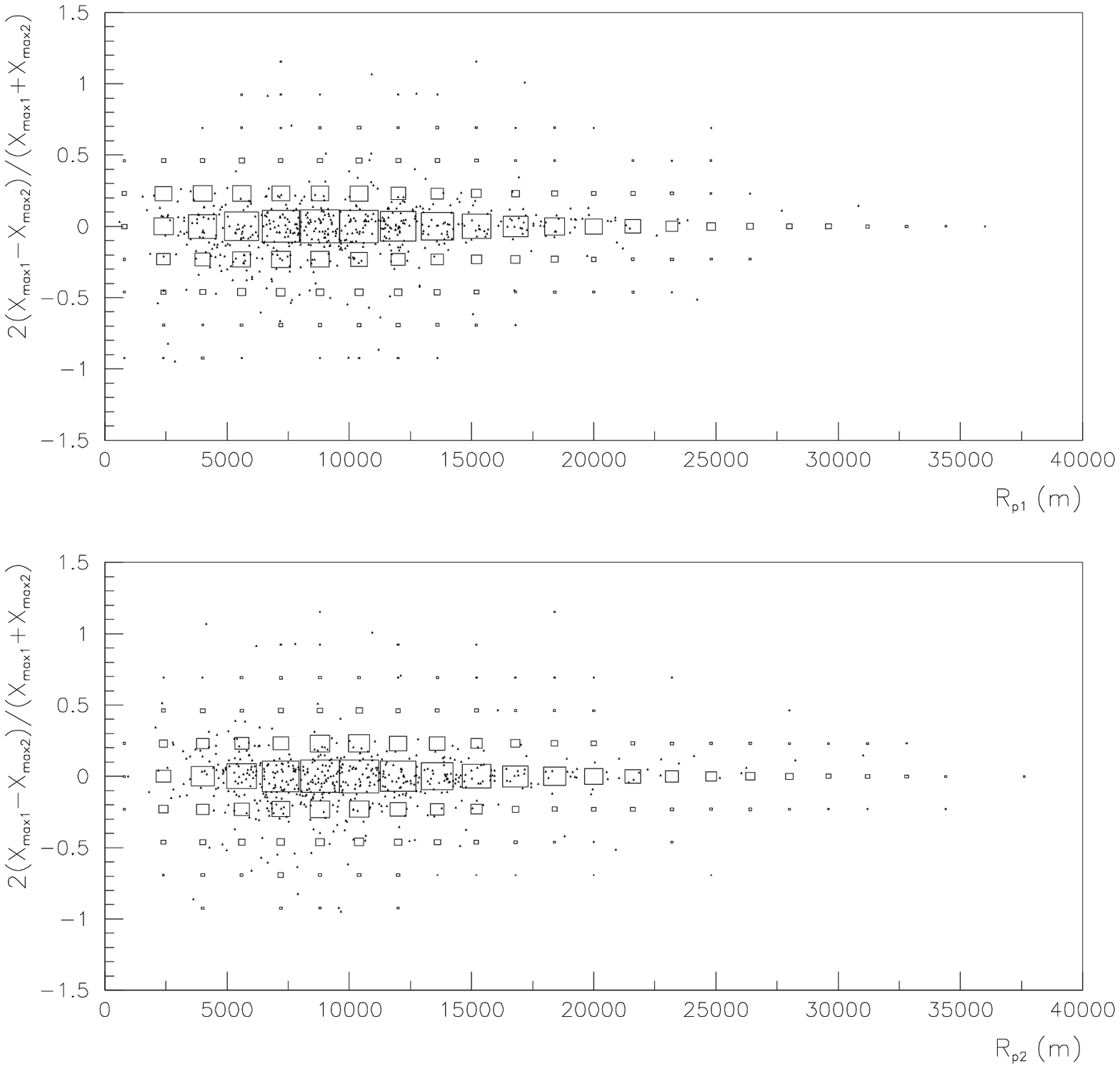}
\caption{Pull as a function of R$_{p}$.  The large boxes represent the Monte Carlo events, and the small triangles represent the data.\label{f17}}
\end{figure}

\begin{figure}
\plotone{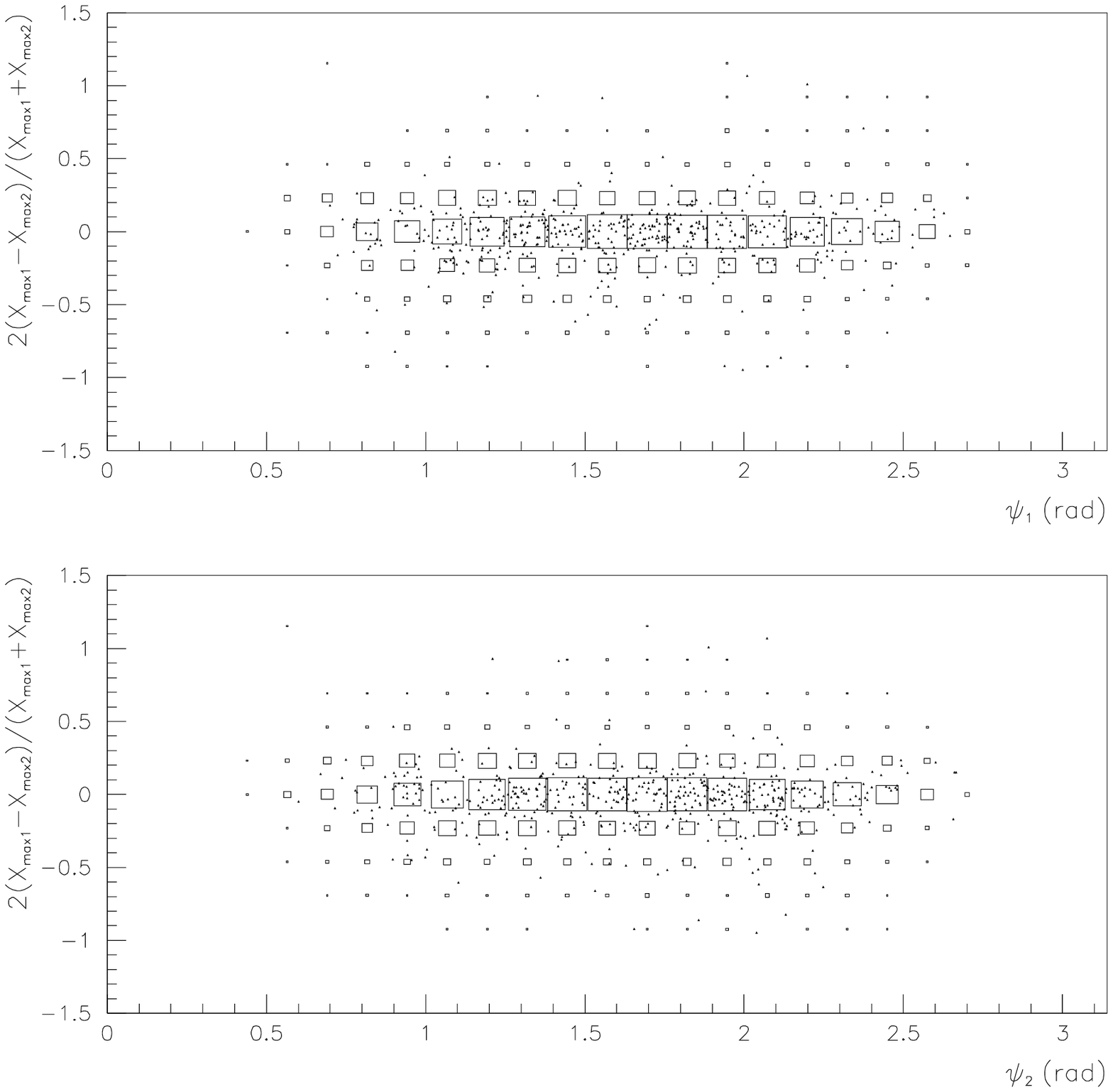}
\caption{Pull as a function of $\psi$.  The large boxes represent the Monte Carlo events, and the small triangles represent the data.\label{f18}}
\end{figure}

Since we will be comparing data and Monte Carlo to extract the cosmic ray composition, it is vital that the Monte Carlo accurately reproduce the detector and reconstruction resolution. This can be demonstrated by examining the differences in the individual detector fits. Figs. \ref{f14} and \ref{f15} show the pull, defined as 2(X$_{max1}$ - X$_{max2}$)/(X$_{max1}$ + X$_{max2}$), where X$_{max1}$ and X$_{max2}$ are from the individual fits by HiRes-1 and HiRes-2, respectively, after all cuts.  The nearly-gaussian shape of the pull and the nearly identical pull distribution for Monte Carlo and data show that the Monte Carlo resolution represents the real detector resolution well.  In addition, Figs. \ref{f16}--\ref{f18} show the dependence of the data and Monte Carlo pull on X$_{max}$, E, R$_p$ and $\psi$ angles.  There is excellent agreement between data and Monte Carlo and no significant dependence of the pull on any of these variables.  While the data pull distribution cannot be directly used to obtain the global X$_{max}$ resolution, the excellent agreement between data and MC pulls implies that the MC estimate of the global X$_{max}$ resolution is reliable.

\section{ELONGATION RATE RESULT}
\label{er_result}

The data were binned in energy as shown in Table \ref{data_table}.
The statistical errors are the standard error of the mean.    

\begin{figure}
\plotone{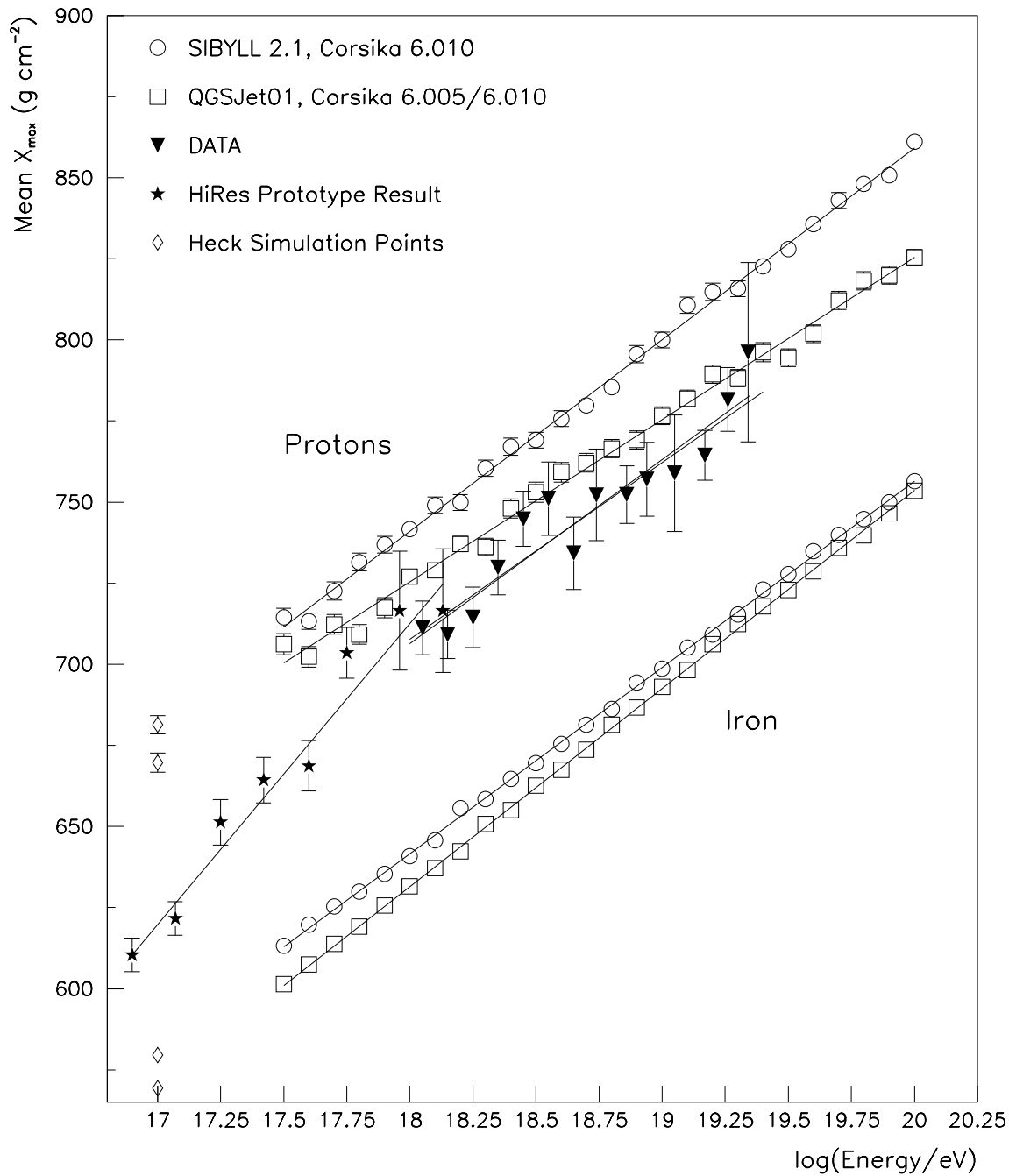}
\caption{Elongation rate result.  The predictions for QGSJet and SIBYLL protons and iron are shown for comparison.  The stars show the HiRes Prototype result.  The diamonds show simulation points calculated by Heck.  The best fit to the data and a fit to the 76\% of the events which have hourly atmospheric corrections are shown.  The latter has a slightly steeper slope.\label{f19}}
\end{figure}

Fig. \ref{f19} shows the ER result.  The QGSJet and SIBYLL
model predictions and the HiRes Prototype result are also indicated.
The measured ER is 54.5 $\pm$ 6.5 g cm$^{-2}$ per decade (statistical uncertainty only; see Section \ref{systematics}), compared to the model predictions of 50
and 61 g cm$^{-2}$ per decade for QGSJet protons and iron nuclei, respectively, and 57 and 59 g cm$^{-2}$ per decade for SIBYLL protons and iron nuclei, as well as to the HiRes Prototype result of 93.0 $\pm$ 8.5 (stat) $\pm$ 10.5 (sys) g cm$^{-2}$ per decade. 

\subsection{Systematic Uncertainty in Elongation Rate}
\label{systematics}

Uncertainties in energy do not have a large affect on ER because of the logarithmic energy scale.  Any systematic uncertainty in X$_{max}$ which applies over the entire energy range will change the absolute value of X$_{max}$, but will not change the ER.  To affect ER, the systematic uncertainty must shift X$_{max}$ in an energy-dependent way.

\begin{figure}
\plotone{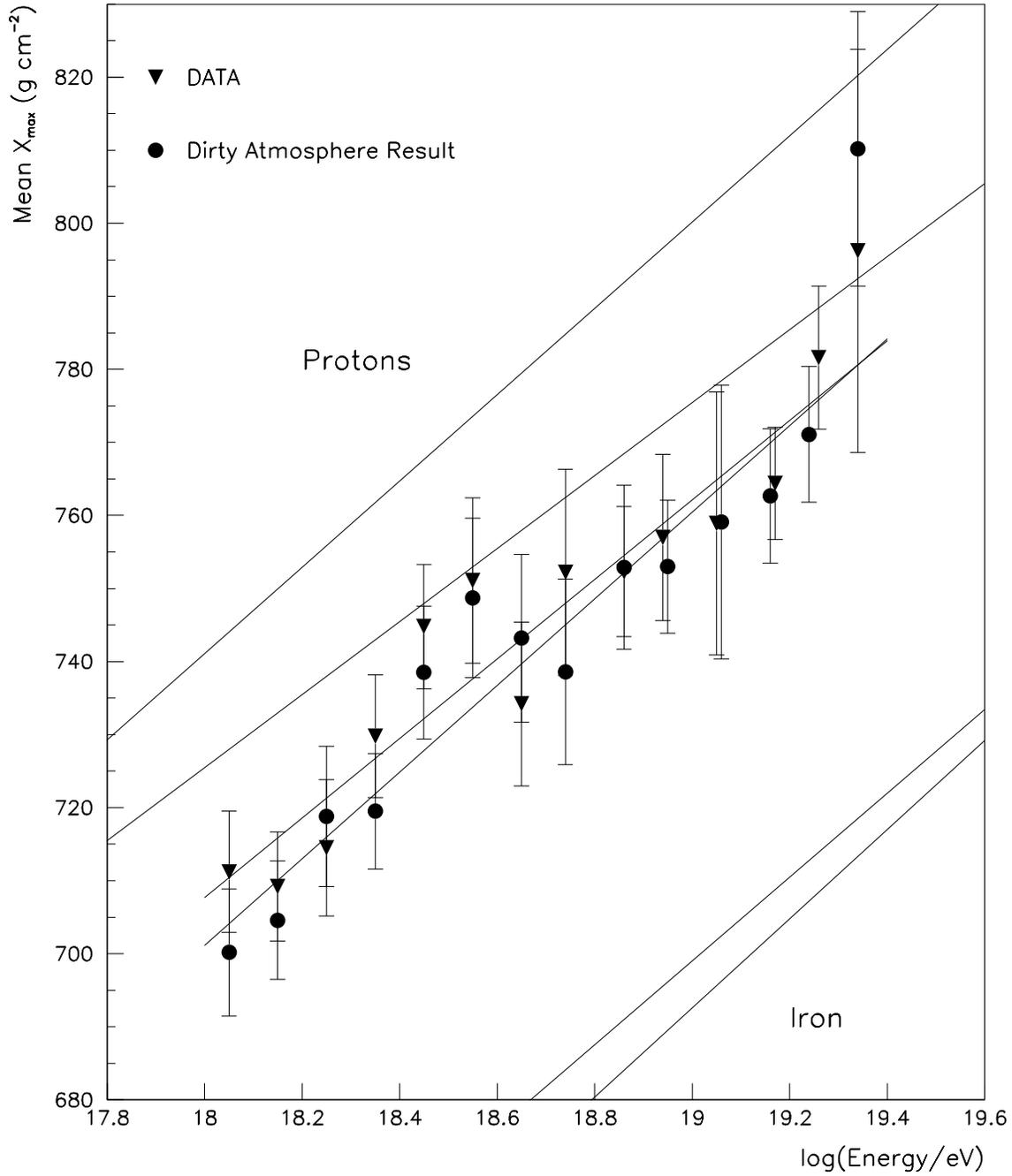}
\caption{Uncertainty in elongation rate.  The circles are obtained by reprocessing the data with a dirtier atmoshpere.  The x- and y-axes are expanded relative to Fig. \ref{f19} to accentuated the small difference.  The upper two and lower two lines are the model predictions.\label{f20}}
\end{figure}

As discussed in Section \ref{atcut}, the measured average atmosphere at HiRes is parameterized by an aerosol vertical  optical depth of 0.04 $\pm$ 0.02 (sys).  Because the database used to obtain this result was used in the reconstruction of about 3/4 of the events, the statistical variation about this mean is already represented in the data and the Monte Carlo.  The effect of the systematic error 
on the mean is studied by re-processing the data with a dirtier atmosphere.  Just as in the original processing, the atmospheric database was sampled and either the database entry or the standard atmosphere was used, as appropriate. Since horizontal extinction is measured separately while the scale height is inferred, for each event, the aerosol scale height was increased such that the optical depth was increased by 0.02.  The dominant effect of the dirtier atmosphere was to increase the reconstructed energies.  The reconstructed X$_{max}$ values also 
decreased slightly, with the two effects combining to steepen the ER, as shown in Fig. \ref{f20}.  Because the energy scale is logarithmic the introduced change in the ER is small. Since 24\% of the data was reconstructed using an average atmosphere, we check that deleting this part of the data does not significantly affect our value for the elongation rate.  This is shown in Fig. \ref{f19}.

The modeling of the \v{C}erenkov beam  in the reconstruction program can introduce an energy-dependent uncertainty in X$_{max}$. While the absolute intensity of \v{C}erenkov radiation is well known, the effective angular distribution in an EAS depends on the multiple scattering of electrons in the atmosphere, which has some uncertainties. To investigate this, the reconstruction code was modified to make the modeled EAS \v{C}erenkov beam 2$^{\circ}$ wider (consistent with one sigma errors in previous measurements), and the data were reprocessed.  The ER was 
essentially unchanged.

\section{X$_{max}$ DISTRIBUTION WIDTH RESULT}
\label{dist_res}

\begin{figure}
\plotone{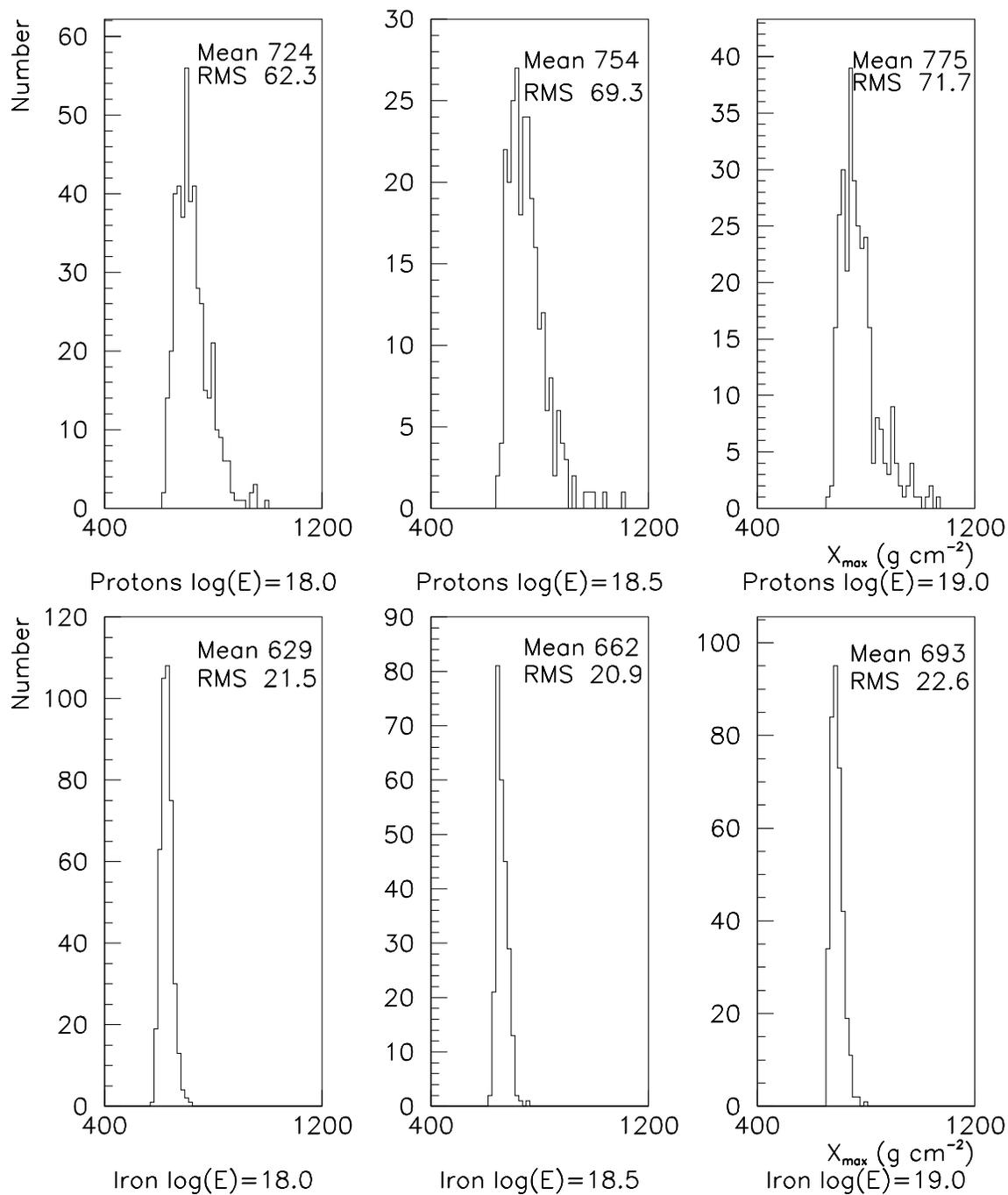}
\caption{X$_{max}$ 
distributions from CORSIKA for primary UHECR energies of 10$^{18.0}$ eV, 10$^{18.5}$ eV, and 10$^{19.0}$ eV.  At each energy, the iron distribution is 
much narrower than the proton distribution.\label{f21}}
\end{figure}

\begin{figure}
\plotone{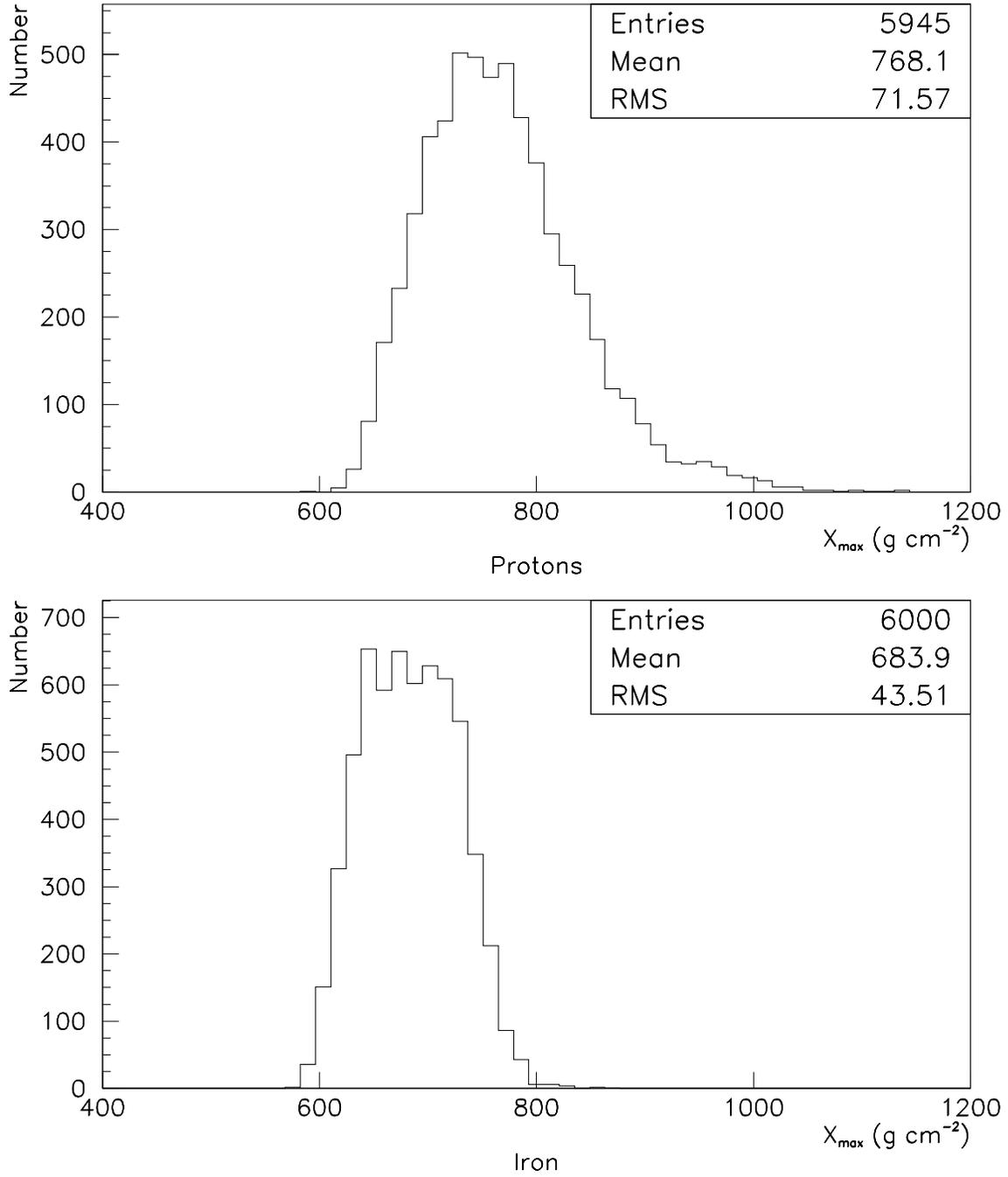}
\caption{ X$_{max}$ distributions from CORSIKA for primary UHECR with energies between 10$^{18.0}$ and 10$^{19.4}$ eV.\label{f22}}
\end{figure}

Protons are expected to show more shower-to-shower fluctuation than iron nuclei.  Fig. \ref{f21} shows that at each of the three energies, both QGSJet and SIBYLL predict that the distribution of X$_{max}$ is wider for proton showers than for iron showers.  Thus if the composition is changing from Fe to protons as energy increases, for example, then the X$_{max}$ distribution at lower energies will be significantly narrower than at higher energies. Fig. \ref{f22} shows 
the X$_{max}$ distributions expected from a purely light or  a purely heavy flux with  a Fly's Eye Stereo spectrum over the entire energy range of interest. For both hadronic interaction models, the difference between iron nuclei and protons is clear even over this large energy range.

\begin{figure}
\plotone{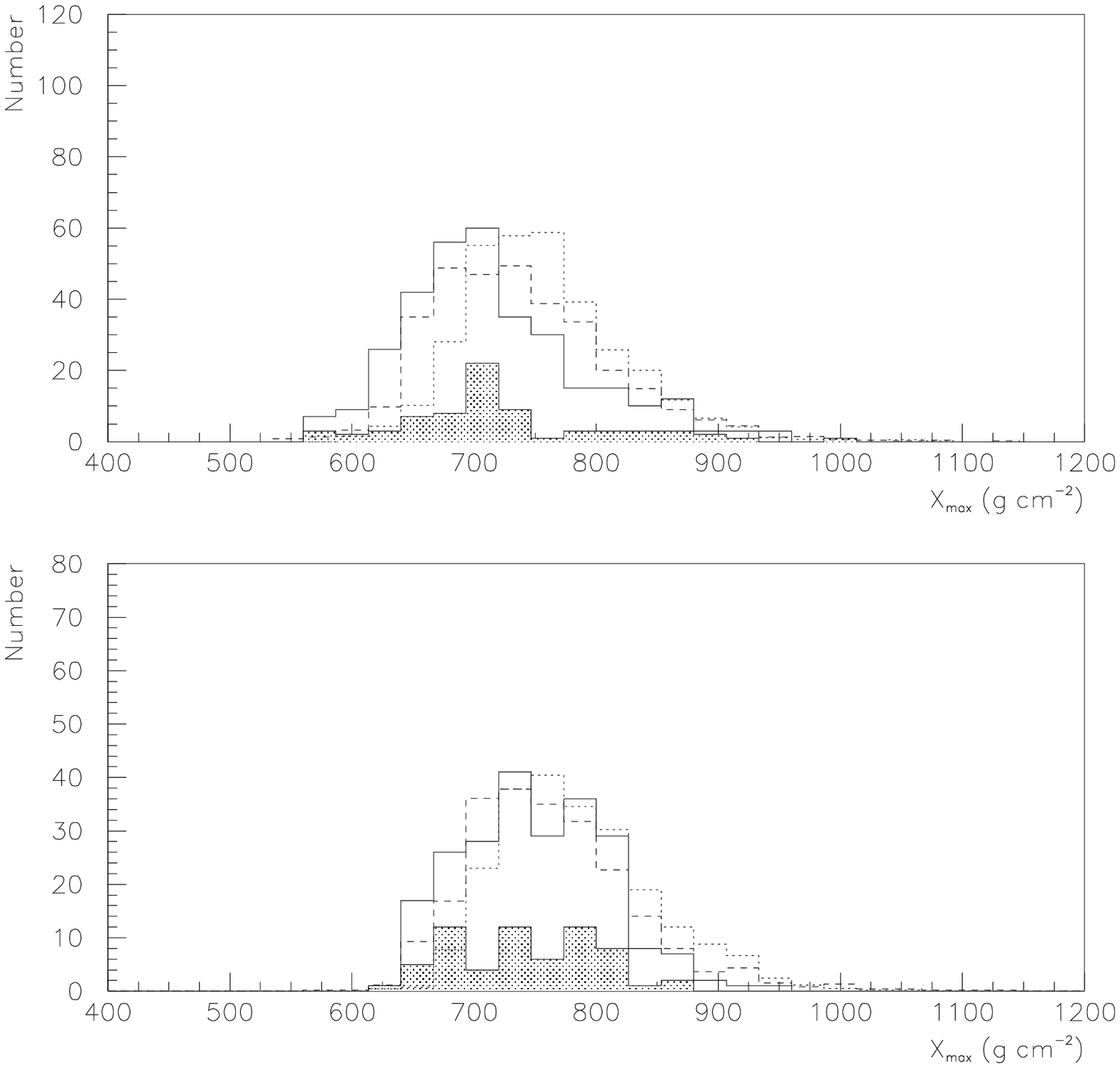}
\caption{Distribution  width results, protons. The top plot is for 
$\log$(E/eV) 18.0 to 18.4 and the bottom plot is for $\log$(E/eV) 
18.4 to 19.4.  In each plot, the solid line is the data, and the shaded area represents the 24\% of the events reconstructed with the average atmosphere.  The dashed line is the QGSJet model, and the dotted line is the SIBYLL model. 
Compare Figs. \ref{f21} and \ref{f24}.\label{f23}}
\end{figure}

\begin{figure}
\plotone{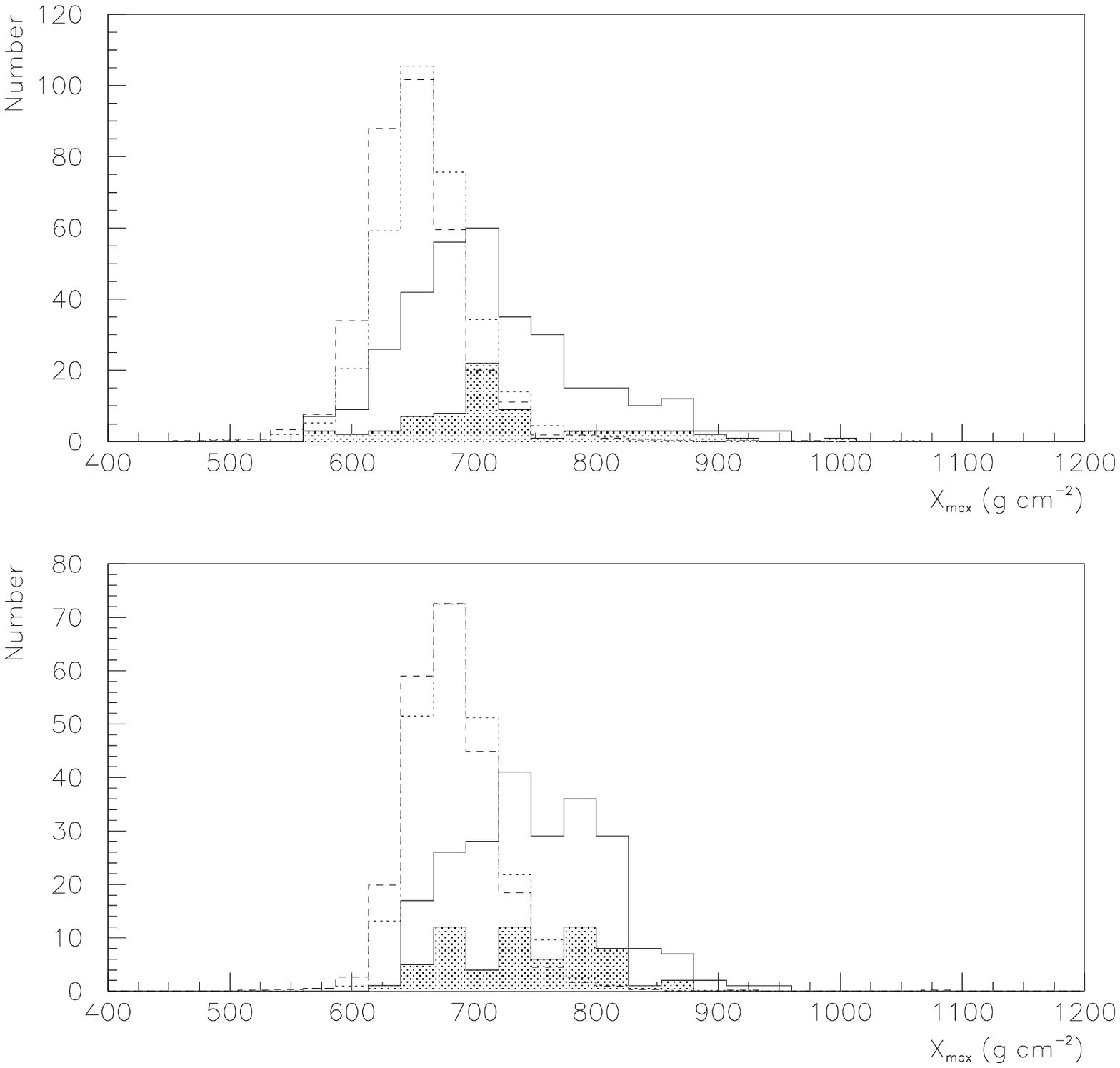}
\caption{Distribution  width results, iron. The top plot is for 
$\log$(E/eV) 18.0 to 18.4 and the bottom plot is for $\log$(E/eV) 
18.4 to 19.4.  In each plot, the solid line is the data, and the shaded area represents the 24\% of the events reconstructed with the average atmosphere.  The dashed line is the QGSJet model, and the dotted line is the SIBYLL model. 
Compare Figs. \ref{f21} and \ref{f23}.\label{f24}}
\end{figure}

\begin{figure}
\plotone{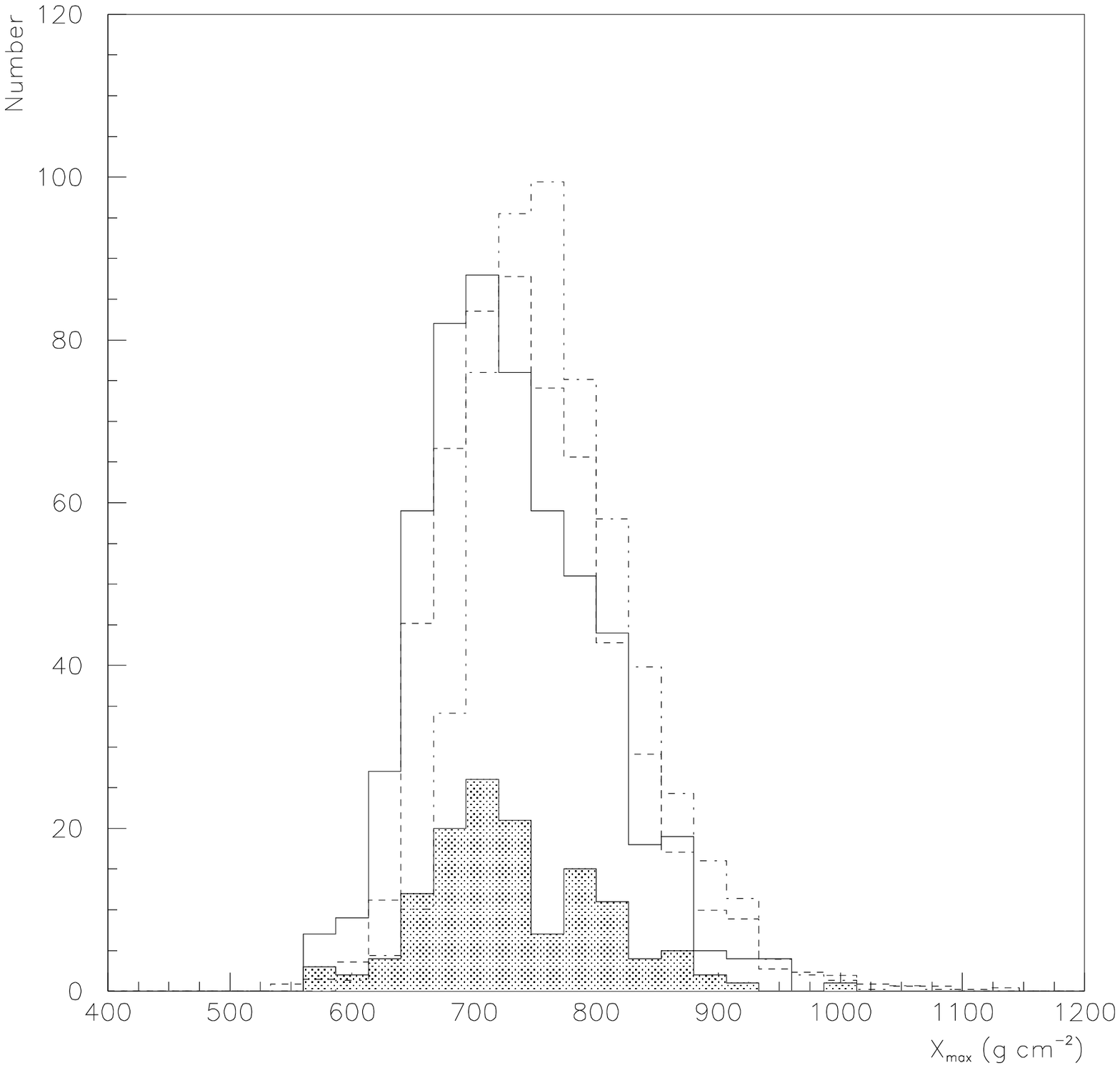}
\caption{All-energy distribution width result, protons.  The solid 
line is the data, and the shaded area represents the 24\% of the events reconstructed with the average atmosphere.  The dashed line is the QGSJet model, and the dotted line is the SIBYLL model.  Compare Figs. \ref{f22} and \ref{f26}.\label{f25}}
\end{figure}

\begin{figure}
\plotone{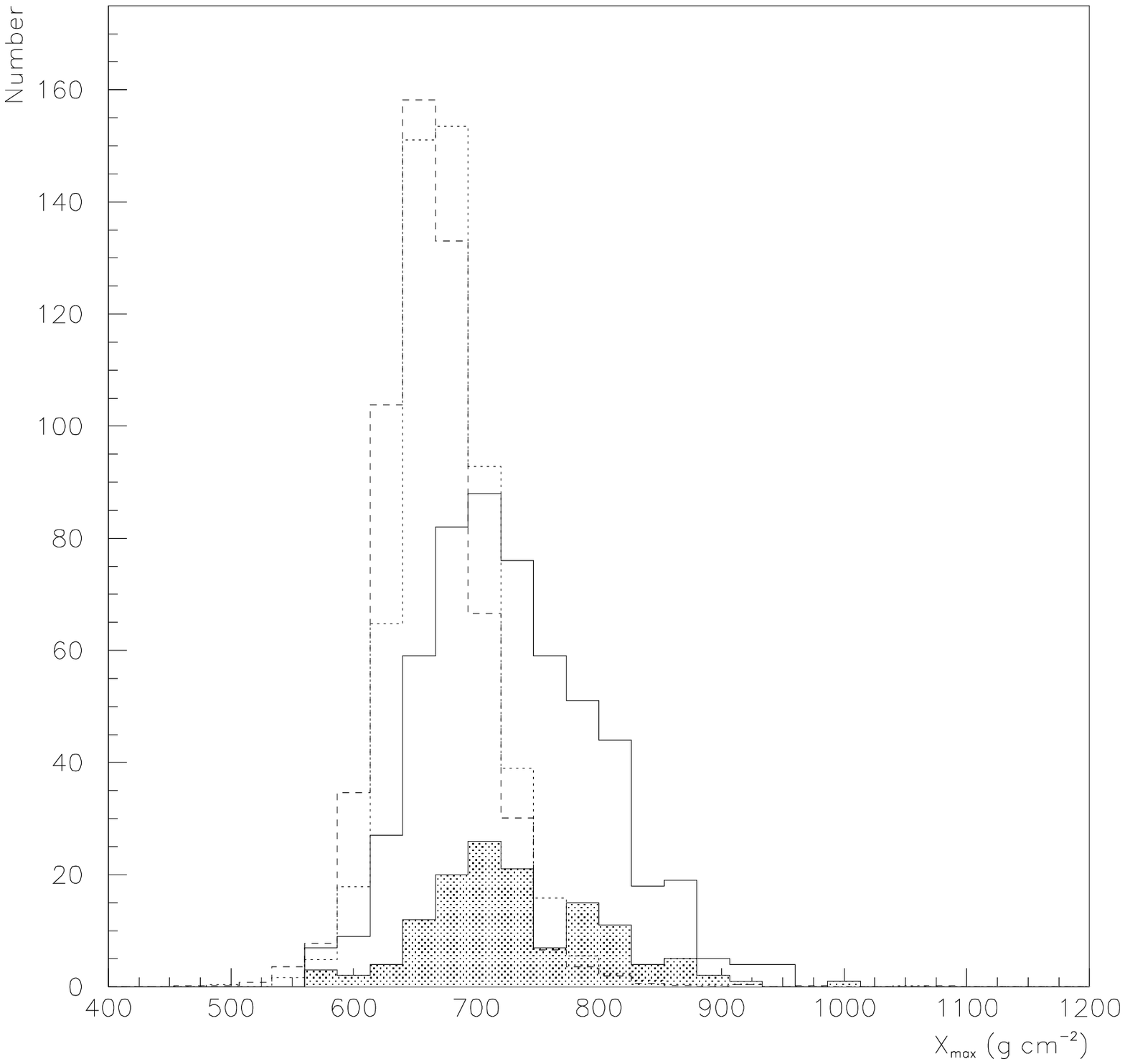}
\caption{All-energy distribution width result, iron.  The solid 
line is the data, and the shaded area represents the 24\% of the events reconstructed with the average atmosphere.  The dashed line is the QGSJet model, and the dotted line is the SIBYLL model.  Compare Figs. \ref{f22} and \ref{f25}.\label{f26}}
\end{figure}

Figs. \ref{f23}--\ref{f26} show the width of the X$_{max}$ distribution. The histograms representing the Monte Carlo in Figs. \ref{f23}--\ref{f26} were obtained by taking over 4500 showers for each model through the complete Monte Carlo and reconstruction routines, subject to the same cuts as the data.  Nearly 2500 events of each type survived.  The areas of the Monte Carlo histograms are normalized to the area of the data histogram.

For Figs. \ref{f23} and \ref{f24}, the data were divided into two energy bins selected such that each contained about half of the events.  The data shows that the width is not changing with energy, indicating that composition is constant or only slowly changing.  The width of the data distribution in Figs. \ref{f25} and \ref{f26} indicates that the composition is predominantly light in agreement with the elongation rate interpretation.  These figures also show the contribution of the 24\% of the data reconstructed with an average atmosphere.  There is no statistically significant difference between this sample and the total data sample.

\begin{figure}
\plottwo{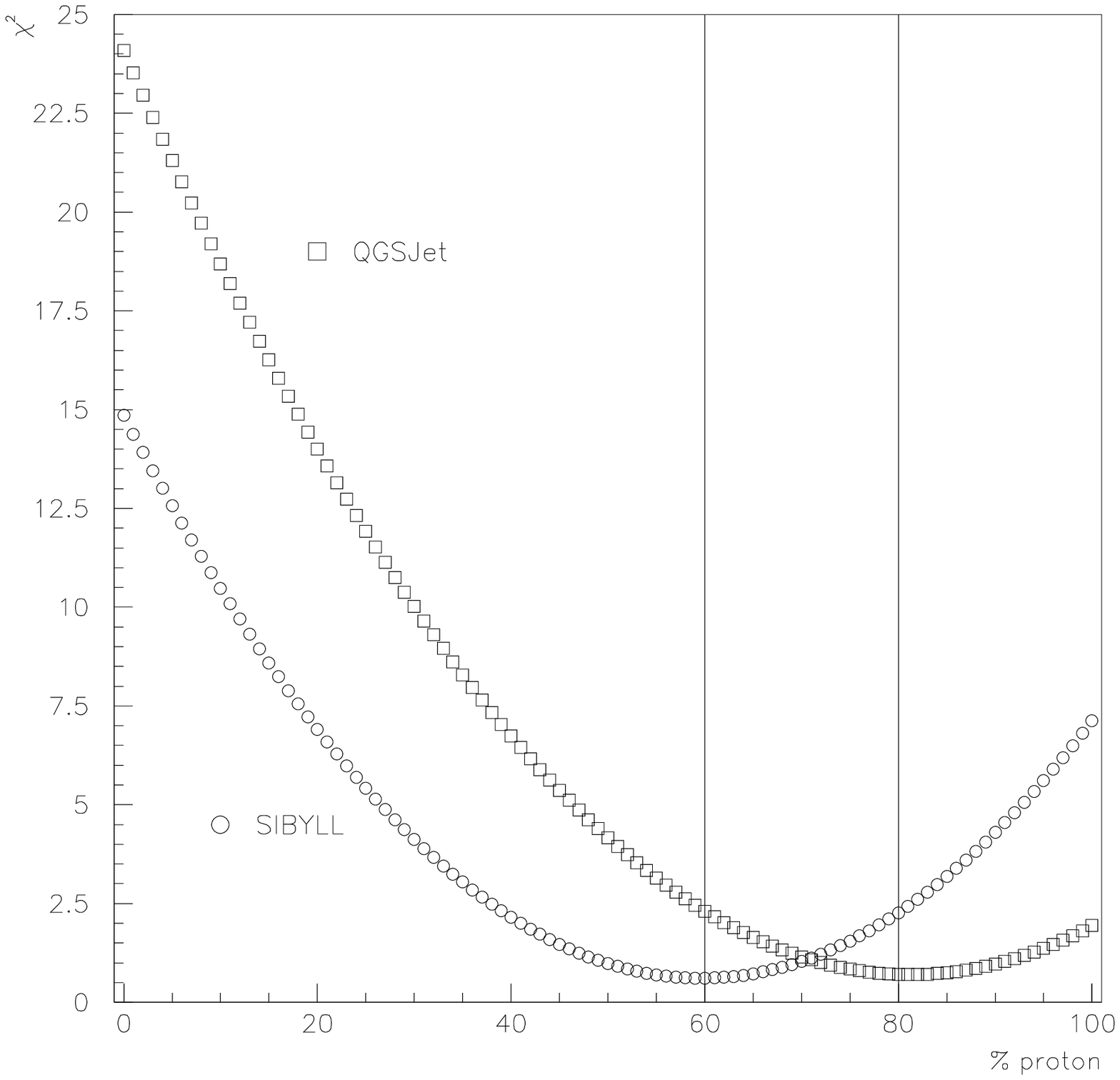}{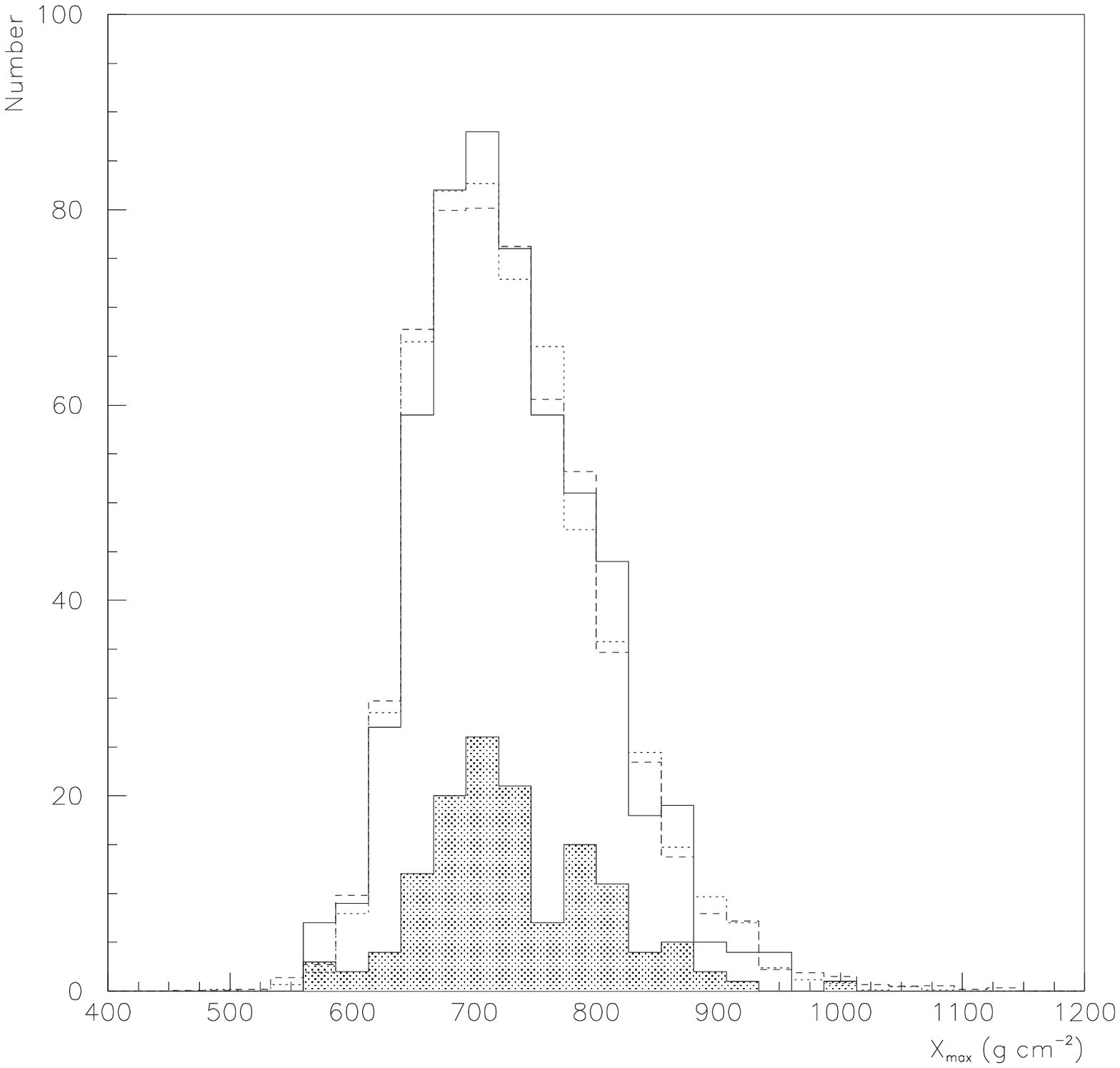}
\caption{Simple two-component composition model fit results.  On the left, the data points 
show the $\chi^2$ for a data/model comparison at each per cent 
proton.  On the right, the solid line is the data, the shaded area represents the 24\% of the events reconstructed with the average atmosphere, the 
dashed line is the QGSJet model with 77\% proton, and the dotted line 
is SIBYLL with 57\% proton.\label{f27}}
\end{figure}

Figs. \ref{f25} and \ref{f26} show that the data are consistent with a nearly purely protonic composition, especially when compared to the QGSJet model.  Assuming a simple two-component model where the primary flux is a mixture of protons and iron nuclei, Fig. \ref{f27} shows how well the model fits the data as a function of fraction of protons.  The best fits are at 80\% protons for QGSJet and 60\% for SIBYLL.  Fig. \ref{f27} also shows the models compared to the data.

\section{SYSTEMATIC UNCERTAINTY IN ABSOLUTE VALUE OF X$_{max}$}
\label{xsys}

Fig. \ref{f27} suggests that the composition of UHECR is predominantly light above $10^{18}$ eV.  However, systematic errors in the absolute value of X$_{max}$ could artificially move the measured X$_{max}$ values too deep in the atmosphere.  The X$_{max}$ values for events with energies above $10^{19}$ eV are of particular interest.  Potential contributors to the systematic uncertainty in X$_{max}$ are biases introduced by reconstruction, errors in PMT pointing directions, variations of the molecular and aerosol component of the atmosphere, and incorrect treatments of the \v{C}erenkov beam.

Resolution studies (Figs. \ref{f2} and \ref{f3})  show that the mean of the reconstructed X$_{max}$ distribution differs by 5 g cm$^{-2}$  from the input value. For events with energy greater than $10^{19}$ eV, the mean of the distribution shifts by 3 g cm$^{-2}$.  The phototube cluster pointing directions have been confirmed by observing stars \citep{bergstar, stefan}. The largest deviation of true pointing direction from the direction used in reconstruction is 0.3$^{\circ}$. This corresponds to a maximum X$_{max}$ error of 15 g cm$^{-2}$. The molecular component of the atmosphere may be different from the U.S. Standard Atmosphere \citep{stdatm} assumed in the reconstruction. The variations in the atmosphere as measured by radiosondes launched from the Salt Lake City airport has been studied \citep{NM}. In the month of the year which varied most from the standard model, the actual pressure differed from the model by 8\%, which leads to  a 10 g cm$^{-2}$ difference in integrated atmospheric depth. Subsequent studies by Y. Fedorova of radiosonde data over several years have confirmed this initial study.  Using an incorrect atmospheric attenuation could distort the shower profile and systematically shift the position of X$_{max}$.  Fig. \ref{f20} shows that above $10^{19}$ eV, an increase as large as 0.02 in VAOD (a 50\% change) does not significantly change the mean values of X$_{max}$. As described in Section \ref{systematics}, the data were reprocessed with a larger value of  \v{C}erenkov beam width.  The mean difference between the X$_{max}$ obtained from reconstruction with the standard \v{C}erenkov beam width and the X$_{max}$ obtained by using a wider beam is less than 1 g cm$^{-2}$ and is  negligible.

Table \ref{sys_table} summarizes the systematic uncertainties in X$_{max}$ for energies above $10^{19}$ eV.  Adding the individual uncertainties in quadrature gives an overall systematic uncertainty of less than 20 g cm$^{-2}$. This is less than the detector resolution and much less than the 75 g cm$^{-2}$ difference between proton and iron mean X$_{max}$.

\section{CONCLUSIONS}
\label{concl}

The measured elongation rate result is consistent with a constant or slowly changing composition between $10^{18.0}$ eV and $10^{19.4}$ eV.  The
data are also in very good agreement with the HiRes Prototype data in
the region where they nearly overlap.  The HiRes Prototype result showed a
composition change from heavy to light in the $10^{17}$ to $10^{18}$
eV range, but the HiRes data do not show a continuation of this
elongation rate. This can be interpreted as strong evidence 
for a transition from a heavy composition to a predominantly light and constant or only slowly changing composition above $10^{18}$ eV.  The earlier Fly's Eye stereo data also shows a transition from a heavy to a light composition though with a broader transition region in energy.  This is consistent with the present data if systematic errors and worse X$_{max}$ resolution quoted in that work are taken into account \citep{new}.

In the present study, the widths of the X$_{max}$ distributions in the UHECR regime strengthen this conclusion.  Such a transition is interesting in light of reported structure in the UHECR spectrum in this same energy region. Many experiments have seen evidence for a second knee in the middle of the $10^{17}$ decade and an ankle structure near 3 x $10^{18}$ eV. The change in composition may reflect a change from a dominant galactic CR flux to an extragalactic flux which dominates near $10^{19}$ eV. Observation of anisotropy from the galactic plane would support this picture.  Some evidence for such anisotropy has been reported \citep{bird99, hay99, bel01}.

\acknowledgements
Parts of this analysis and the preparation of this document were performed at Lawrence Livermore National Laboratory, which is operated by the University of California for the Department of Energy. One of us (Sokolsky) would like to thank the Guggenheim Foundation for support. The HiRes project is supported by the NSF under contract numbers NSF-PHY-9322298, NSF-PHY-9904048, NSF-PHY-9974537, NSF-PHY-0098826, NSF-PHY-0305516 and NSF-PHY-0245428, by the Department of Energy Grant FG03-92ER40732, and by the Australian Research Council.  The cooperation of Colonel E. Fisher, US Army Dugway 
Proving Ground, and his staff is appreciated.

\clearpage


\clearpage

\clearpage

\clearpage

\clearpage

\clearpage

\clearpage

\clearpage

\clearpage

\clearpage

\clearpage

\clearpage

\clearpage

\clearpage

\clearpage

\clearpage

\clearpage

\clearpage

\clearpage

\clearpage

\clearpage

\clearpage

\clearpage

\clearpage

\clearpage

\clearpage

\clearpage

\clearpage


\begin{deluxetable}{l c}
\tablecaption{Quality cuts.\label{cut_table}}
\tablehead{\colhead{Parameter} & \colhead{Cut}}
\startdata
Minimum Viewing Angle (both sites) & 10\\
Minimum Opening Angle Between SDPs & 5\\
Individual Site Fit $\chi^2$ & 20\\
Global Fit $\chi^2$ & 15\\
Timing Fit X$_{max}$ - Global Fit X$_{max}$ & 500\\
HiRes-1 Fit X$_{max}$ - HiRes-2 Fit X$_{max}$ & 500\\
Geometric Uncertainty in Individual Site Fit & 400\\
Geometric Uncertainty in Global Fit & 200\\
Bracketing X$_{max}$ & X$_{max}$ within 60 of viewed track\\
\enddata
\tablecomments{Angles are in degrees, $\chi^2$ are unitless, and X$_{max}$ are in g cm$^{-2}$.}
\end{deluxetable}

\clearpage

\begin{table}
\begin{center}
\caption{Elongation rate data.\label{data_table}}
\begin{tabular}{c c c r @{ $\pm$ } r @{.} l}
\tableline\tableline
$\log$(E/eV) & Number of & Mean &
\multicolumn{3}{c}{Mean X$_{max}$} \\
Bin& Events &$\log$(E/eV)&\multicolumn{3}{c}{(g cm$^{-2}$)} \\
\tableline
18.0-18.1&92&18.05 $\pm$ 0.01&711.3&8&2\\
18.1-18.2&79&18.15 $\pm$ 0.01&709.2&7&5\\
18.2-18.3&82&18.25 $\pm$ 0.01&714.5&9&3\\
18.3-18.4&74&18.35 $\pm$ 0.01&729.8&8&4\\
18.4-18.5&59&18.45 $\pm$ 0.01&744.8&8&5\\
18.5-18.6&34&18.55 $\pm$ 0.01&751.1&11&3\\
18.6-18.7&29&18.65 $\pm$ 0.01&734.2&11&2\\
18.7-18.8&23&18.74 $\pm$ 0.01&752.2&14&1\\
18.8-18.9&23&18.86 $\pm$ 0.01&752.3&8&9\\
18.9-19.0&20&18.94 $\pm$ 0.01&757.0&11&4\\
19.0-19.1&13&19.05 $\pm$ 0.01&758.9&18&0\\
19.1-19.2&14&19.17 $\pm$ 0.01&764.4&7&7\\
19.2-19.3&6&19.26 $\pm$ 0.01&781.6&9&8\\
19.3-19.4&5&19.35 $\pm$ 0.01&796.2&27&7\\
\tableline
\end{tabular}
\tablecomments{Uncertainties are the standard error of the means.}
\end{center}
\end{table}

\clearpage

\begin{deluxetable}{l c}
\tablecaption{ Systematic uncertainties in X$_{max}$.\label{sys_table}}
\tablehead{\colhead{Uncertainty} & \colhead{g cm$^{-2}$}}
\startdata
Pointing Direction&15\\
Atmospheric Variations&10\\
Reconstruction Bias&5\\
Sum in Quadrature&18.7\\
\enddata
\end{deluxetable}

\end{document}